\documentclass[10pt, journal, doublecolumn]{IEEEtran}

\ifCLASSINFOpdf
  \usepackage[pdftex]{graphicx}
   \graphicspath{{../figures/}{../jpeg/}}
\else
\fi

\usepackage[cmex10]{amsmath}
\usepackage{amssymb}
\usepackage{bm}

\usepackage{algpseudocode}
\usepackage{algorithm}
\usepackage{array}
\usepackage{url}
\usepackage[noadjust]{cite}

\newtheorem{remark}{Remark}

\usepackage{fancyhdr}
\usepackage{amsfonts}
\usepackage{xfrac}
\usepackage{color}
\usepackage{pgfplots}           
\pgfplotsset{compat=1.16}
\usepackage{enumerate}

\usepackage{cite}
\usepackage{mathrsfs}
\usepackage{subcaption}
\usepackage{stfloats}

\usepackage{afterpage}
\usepackage{tikz}
\usepackage{mathdots}
\usepackage{yhmath}
\usepackage{cancel}
\usepackage{color}
\usepackage{siunitx}
\usepackage{array}
\usepackage{multirow}
\usepackage{textcomp}
\usepackage{gensymb}
\usepackage{tabularx}
\usepackage{booktabs}
\usepackage{graphicx}
\usepackage{xcolor}

\allowdisplaybreaks

\usetikzlibrary{decorations.markings,decorations.pathmorphing}
\begin{document}

\title{Energy-aware Trajectory Optimization for UAV-mounted RIS and Full-duplex Relay}
\author{Dimitrios Tyrovolas,~\IEEEmembership{Graduate Student Member,~IEEE,}
Nikos A. Mitsiou,~\IEEEmembership{Student Member,~IEEE,} \\Thomas G. Boufikos, Prodromos-Vasileios Mekikis,~\IEEEmembership{Member,~IEEE,} Sotiris A. Tegos,~\IEEEmembership{Member,~IEEE,} \\
Panagiotis D. Diamantoulakis,~\IEEEmembership{Senior Member,~IEEE,} Sotiris Ioannidis, Christos K. Liaskos,~\IEEEmembership{Member,~IEEE,} \\ and George K. Karagiannidis,~\IEEEmembership{Fellow,~IEEE}        \thanks{D. Tyrovolas is with the Dept. of Electrical and Computer Engineering, Aristotle University of Thessaloniki, 54124 Thessaloniki, Greece, and with the Dept. of Electrical and Computer Engineering, Technical University of Crete, Chania, Greece (tyrovolas@auth.gr).}  
\thanks{N. A. Mitsiou, T. G. Boufikos, S. A. Tegos, and P. D. Diamantoulakis are with the Dept. of Electrical and Computer Engineering, Aristotle University of Thessaloniki, 54124 Thessaloniki, Greece (\{nmitsiou, mpothogeo, tegosoti, padiaman\}@auth.gr).} 
\thanks{P.-V. Mekikis is with Hilti Corporation, Feldkircher Strasse 100, 9494 Schaan, Liechtenstein (akis.mekikis@hilti.com)}
\thanks{S. Ioannidis is with the Dept. of Electrical and Computer Engineering, Technical University of Crete, Chania, Greece (sotiris@ece.tuc.gr).}
\thanks{C. K. Liaskos is with the Computer Science Engineering Department, University of Ioannina, Ioannina, and with the Foundation for Research and Technology Hellas (FORTH), Greece (cliaskos@ics.forth.gr).} 
\thanks{G. K. Karagiannidis is with the Dept. of Electrical and Computer Engineering, Aristotle University of Thessaloniki, 54124 Thessaloniki, Greece and with the Artificial Intelligence \& Cyber Systems Research Center, Lebanese American University (LAU), Lebanon (geokarag@auth.gr).}
\thanks{The work has been funded by the European Union’s Horizon 2020 research and innovation programs under grant agreement No 101021659 (SENTINEL).}
}

\maketitle

\begin{abstract}
In the evolving landscape of sixth-generation (6G) wireless networks, unmanned aerial vehicles (UAVs) have emerged as transformative tools for dynamic and adaptive connectivity. However, dynamically adjusting their position to offer favorable communication channels introduces operational challenges in terms of energy consumption, especially when integrating advanced communication technologies like reconfigurable intelligent surfaces (RISs) and full-duplex relays (FDRs). To this end,
by recognizing the pivotal role of UAV mobility, the paper introduces an energy-aware trajectory design for UAV-mounted RISs and UAV-mounted FDRs using the decode-and-forward (DF) protocol, aiming to maximize the network's minimum rate and enhance user fairness, while taking into consideration the available on-board energy. Specifically, this work highlights their distinct energy consumption characteristics and their associated integration challenges by developing appropriate energy consumption models for both UAV-mounted RISs and FDRs that capture the intricate relationship between key factors such as weight, and their operational characteristics. Furthermore, a joint time-division multiple access (TDMA) user scheduling-UAV trajectory optimization problem is formulated, considering the power dynamics of both systems, while assuring that the UAV energy is not depleted mid-air. Finally, simulation results underscore the importance of energy considerations in determining the optimal trajectory and scheduling and provide insights into the performance comparison of UAV-mounted RISs and FDRs in UAV-assisted wireless networks.

\end{abstract}

%
\IEEEpeerreviewmaketitle

\vspace{-0.3cm}
\section{Introduction}

Unmanned aerial vehicles (UAVs) are emerging as transformative tools in the landscape of future sixth-generation (6G) wireless networks \cite{zorzi2021,Mekikis2023}. Specifically, their inherent flexibility allows them to follow optimized trajectories, dynamically adjusting their paths to offer line-of-sight (LoS) channels ubiquitously. This capability makes them invaluable in scenarios demanding adaptive connectivity solutions, such as bridging connectivity gaps in challenging terrains or enhancing network resilience in disaster-stricken areas \cite{zorzi2021,Matracia2023}. Furthermore, their ability to provide on-demand high-capacity coverage in crowded events or remote locations underscores their pivotal role in reshaping wireless communication. However, despite their numerous advantages, UAVs face the challenge of limited onboard energy that dictates their operational time \cite{ruizhang2017, YongZeng2019}, underscoring the need to optimize their operational efficiency. To this end, innovative methods are essential to harness the full potential of UAVs, while taking into account their energy constraints.

Given the finite flight duration of UAVs, it becomes of paramount importance to maximize data throughput and efficiency during their operational time. In this direction, full-duplex (FD) communication, with its ability to simultaneously transmit and receive data, emerges as a key solution in this context \cite{hanzo2016}. Unlike traditional half-duplex systems that alternate between transmission and reception, FD systems effectively double the spectral efficiency, making every time slot of the UAV's flight time count \cite{hanzo2016,shende2018}. This capability is particularly beneficial for UAVs, which can dynamically adjust their positions to establish LoS communication links, without the constraints of ground-based systems. To realize the potential of FD communications in UAVs, the two promising technologies that come to the forefront are \textit{reconfigurable intelligent surfaces (RISs)} and \textit{full-duplex relays (FDRs)} \cite{yuanweiliu2021,linzhang2017}. Specifically, an RIS is an advanced technology that can be implemented as either a programmable reflectarray or a programmable metasurface, both designed to manipulate electromagnetic (EM) waves for enhanced communication. Specifically, programmable reflectarrays offer dynamic backscattering and phase shifting of incident waveforms through omnidirectional antennas with controllable termination, while programmable metasurfaces extend these capabilities to include anomalous reflection angles and polarization manipulation \cite{reflectarrays2020}. This capability allows RIS to support FD communications by dynamically adjusting signal paths with minimal power requirements, significantly improving connectivity and signal quality \cite{liaskos2018,basar2019,direnzo2020}. On the other hand, FDRs enhance spectral efficiency by enabling simultaneous transmission and reception of signals over the same frequency channel, using distinct sets of antennas for each task \cite{shende2018,gangliu2015}. In addition, unlike RISs, the operation of FDRs is also based on the decode-and-forward (DF) protocol, which involves the decoding of the received message before forwarding, ensuring transmission of only accurately decoded messages. Therefore, as the integration of these technologies with UAVs continues to evolve, understanding their distinct advantages and challenges becomes crucial in shaping the future of UAV-assisted wireless networks.

\subsection{Related Works}
In recent years, the integration of UAVs with RIS and FDRs has gathered significant attention in the research community. Specifically, numerous studies have delved into the intricacies of UAV-mounted RIS and UAV-mounted FDRs, exploring various aspects ranging from performance optimization to energy efficiency, and underscoring their potential to reshape the dynamics of wireless communication.

\subsubsection{UAV-mounted RIS}
    As researchers explore ways to optimize wireless communications, the integration of UAVs with RISs emerges as a promising solution to maintain LoS links, especially in propagation environments where the wireless links can often be obstructed \cite{yihong2022}. Specifically, the authors in \cite{haas2021} showcased that UAV-mounted RIS can improve outage performance in dense urban scenarios, even with the dynamic mobility of UAVs. Additionally, \cite{trung2021} explored strategies for the optimal deployment of UAV-mounted RIS in URLLC systems, focusing on scenarios where user fairness is of paramount importance, while \cite{chatzinotas2022} proposed a novel system design that leverages ambient backscatter communication in UAV-mounted RIS networks. Interestingly, considering the different electromagnetic functionalities of RIS \cite{liaskos2018}, the authors in \cite{pitilakis2023} provided a rigorous path loss model for the case where a UAV-mounted absorbing metasurface is utilized and validated the findings experimentally in an anechoic chamber. Transitioning to trajectory design, \cite{haibomei2022} emphasized on three-dimensional (3D) trajectory design for UAVs in urban environments, aiming to optimize the signal-to-noise ratio (SNR) for ground users, while \cite{huilong2020} presented a trajectory optimization framework for UAV-mounted RIS focusing on maximizing the network's secure energy efficiency. Finally, \cite{binduo2023} examined the joint optimization of the UAV trajectory design and the RIS design to facilitate the offloading of computational tasks in IoT networks. To this end, it is imperative to consider both communication and trajectory design for realizing the full potential of UAV-mounted RIS in diverse wireless network scenarios.
\subsubsection{UAV-mounted FDR}
While UAV-mounted RISs offer unique advantages in wireless communications, they are often challenged by significant path loss due to the double path loss phenomenon \cite{basar2019}. In contrast, UAV-mounted FDRs, equipped with integrated electronic components such as amplifiers and decoders, not only have the capability to transmit and receive signals simultaneously but can also process the received signal, making them more robust to path loss than RISs. Therefore, UAV-mounted FDRs present a compelling solution, especially in dynamic environments demanding real-time data exchange and reduced latency \cite{shende2018,linzhang2017}. By taking into account the advantages of FDRs, \cite{menghua2018} delved into optimizing the source and the UAV-mounted FDR transmit power along with its trajectory to enhance the system's outage probability. Furthermore, the potential of UAV-mounted FDRs in high-frequency scenarios was highlighted by \cite{depaiva2021} and \cite{lipengzhu2020}, with a focus on millimeter-wave channels. Moreover, considering the increasing importance of secure communications, \cite{binduo2020} introduced a secrecy communication scheme using a UAV-mounted FDR, optimizing various parameters to ensure both energy efficiency and security. Lastly, \cite{bingli2021}, and \cite{WeiWang2022} proposed an optimization algorithm for the UAV trajectory, user scheduling, and FDR power, showcasing significant performance improvements in scenarios with multiple ground users. Therefore, these recent research advances underscore the versatility of UAV-mounted FDRs in addressing diverse communication challenges in modern wireless networks.

\subsection{Motivation \& Contribution}
In light of the aforementioned works, both RIS and FDR have been extensively compared to determine their respective advantages.
However, a critical oversight concerning comprehensive energy consumption models in the analysis of UAV-mounted systems remains prevalent in the existing literature. For instance, while nearly passive, an RIS can become considerably larger due to the numerous reflecting elements, adding weight and impacting the UAV's energy efficiency \cite{tyrovolas2023}. Conversely, an FDR, though lightweight, demands more energy for tasks like decoding, amplification, and self-interference (SI) mitigation \cite{lipengzhu2020,shaikh2021}. Additionally, the different path loss characteristics of these systems introduce uncertainty in the UAV energy consumption for traversing, as the UAV may need to navigate to various locations to optimize path loss, further complicating the UAV energy consumption. To this end, neglecting detailed energy consumption models in trajectory optimization may lead to incomplete or imprecise assessments of the operational capabilities and limitations of UAV-mounted RIS and FDR systems. Building on this, few studies have focused on the performance of UAV-mounted systems, while considering their energy consumption profiles. For instance, \cite{tyrovolas2023} and \cite{yue2023} explored UAV-mounted RIS-based communications, emphasizing the RIS weight in UAV energy consumption and identifying an optimal number of reflecting elements, with the latter adjusting this number with the addition of a solar panel. Finally, even though \cite{mdpi2023} compared UAV-mounted RIS with their relay counterparts, it overlooked the energy consumption characteristics and trajectory planning, thus potentially leading to incomplete conclusions. Therefore, to the best of the authors' knowledge, no existing work provides a comprehensive comparison of UAV-mounted RIS and UAV-mounted FDRs, especially in the context of optimal trajectory design, while considering their distinct energy consumption profiles.

In this paper, a comprehensive analysis of both UAV-mounted RIS and UAV-mounted FDR employing the DF protocol is presented. In more detail, our contribution is the following:
\begin{itemize}
    \item  We devise appropriate energy consumption models for both UAV-mounted RIS and UAV-mounted FDR that accurately capture the intricate relationship between key factors such as weight, flight duration, and the operational needs of RISs and FDRs in terms of energy.
    \item  Recognizing the intricacies of UAV-mounted RIS and FDR setups, we formulate a joint time division multiple access (TDMA) user scheduling and UAV trajectory optimization problem that accounts for the power dynamics associated with both technologies. Given the non-convex nature of this optimization problem, we employ a combination of alternate optimization and successive convex optimization techniques, ensuring an efficient approach to obtaining an approximate optimal solution.
    \item Through simulation results, we demonstrate how our proposed methods significantly enhance the network minimum rate and user fairness. More specifically, our results show that for UAV-mounted RIS, increasing the number of reflecting elements does not necessarily translate into improved performance, largely due to the added weight of a larger RIS that limits operational flight time. In a notable shift from existing assumptions, the UAV-mounted FDR consistently outperforms the nearly passive RIS, which underscores the key role of UAV motors and the associated weight in overall UAV energy consumption. Additionally, the results highlight the crucial role of the UAV's battery capacity in trajectory optimization, directly influencing the optimal trajectory and thereby necessitating UAV movement only when it is essential for minimizing energy consumption during traversal. To this end, our work emphasizes the importance of energy-aware design in UAV-assisted communication networks, focusing on balancing energy consumption with communication efficiency.
\end{itemize}

\subsection{Structure}
The remaining of the paper is organized as follows. The system model is described in Section \ref{section:system model}. Furthermore, the examined optimization problem and its solution is presented in Section \ref{section:energy}, while our simulation results are presented in Section \ref{section:simulation}. Finally, Section \ref{section:conc} concludes the paper.

\section{System Model}\label{section:system model}
\subsection{System Overview}

\begin{figure}[!t]
	\centering
\includegraphics[width=0.65\columnwidth]{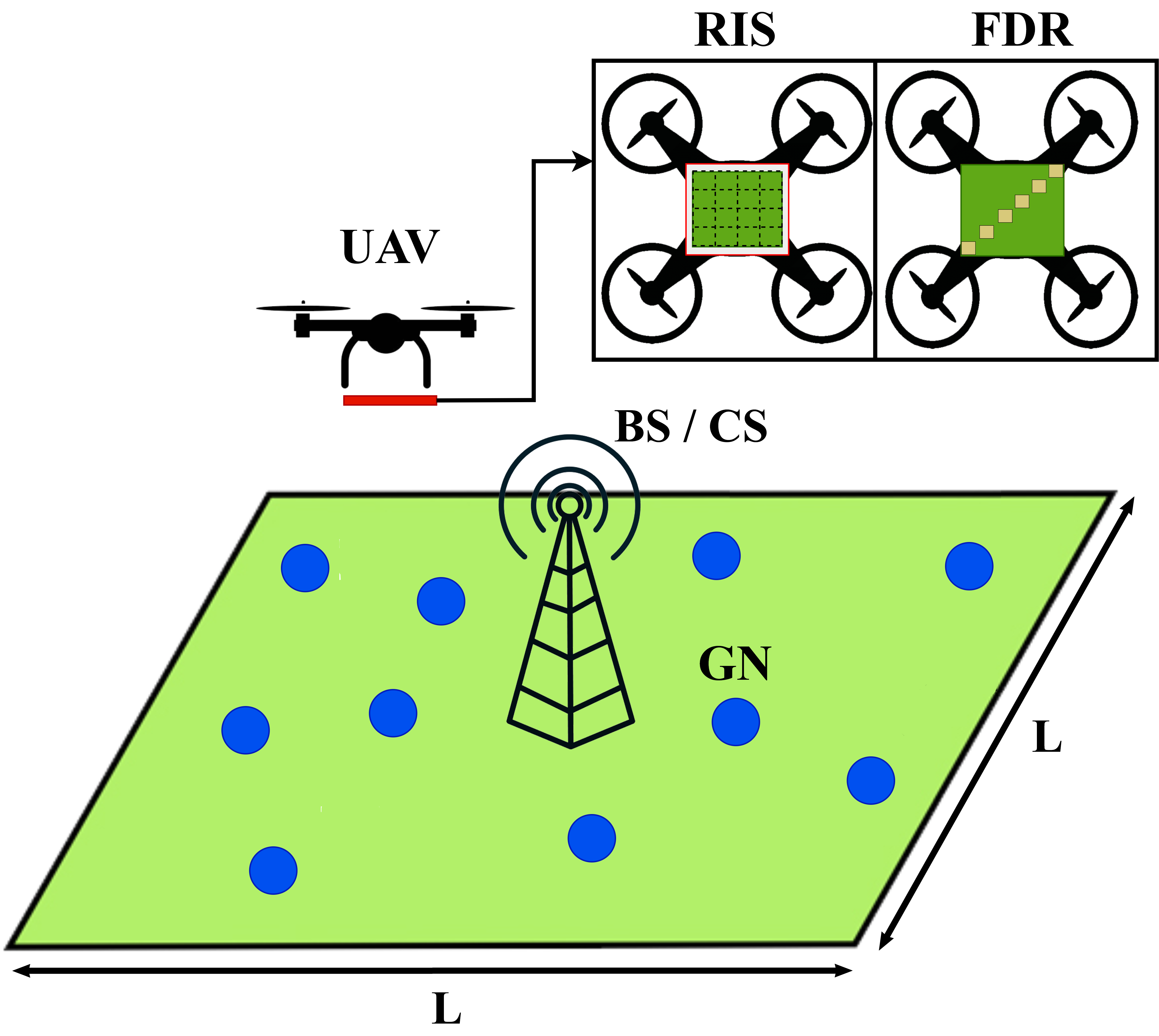}
	\caption{UAV-assisted network topology.}
	\label{sysmod_fig}
\end{figure}

We examine a network of $K$ ground nodes (GNs) that are randomly distributed over a rectangular region with sides equal to $L$, and a base station (BS), which also operates as a UAV charging station (CS). However, given the challenging propagation conditions due to excessive distances, physical obstructions like buildings, and the GNs' limited transmission power, it is assumed that direct communication links between each GN and the BS are not available. To address this, we employ a rotary-wing UAV equipped with either an RIS or an FDR, to act as an intermediate assisting node between the GNs and the BS. Specifically, the UAV takes off from the BS and establishes LoS communication between the GNs and the BS, while flying along a designated trajectory. Afterwards, the UAV returns to the BS for recharging purposes, leveraging the BS's dual functionality as a CS \cite{mekikis2019}. It should be highlighted that the examined network employs TDMA, a channel access method that allocates distinct time slots to multiple users within the same frequency band. This approach allows multiple GNs to share the same communication channel without interference, ensuring efficient service to different ground nodes within the UAV's flight duration. Finally, as illustrated in Fig. 1, it is crucial to ensure that the communication equipment is appropriately attached to the UAV frame to avoid disrupting the airflow around the motors and, thus, compromising its aerodynamics and stability \cite{yue2023}.

Considering a 3D Cartesian coordinate system, we assume that the rectangular region's center coincides with the origin of the coordinate system, the BS location is equal to $\boldsymbol{l}_{\mathrm{\mathbf{BS}}}=\left[0, 0, H_{\mathrm{BS}}\right]$, where $H_{\mathrm{BS}}$ represents the BS height, and the $K$ GNs are located at $\boldsymbol{l}_{\mathrm{\mathbf{k}}}=[x_k, y_k, 0]$, where $k \in \{1,\ldots, K\}$, respectively. Additionally, we assume that the UAV flies at a fixed altitude $H_u$, which is selected appropriately to ensure that the UAV navigates clear of any environmental obstacles. Furthermore, considering that the trajectory duration equals to $T$, the UAV location at time $t$ can be written as $\boldsymbol{q}{\left(t\right)}=[x_{q}(t), y_{q}(t), H_u]$,  where $0\leq t \leq T$ and $\left(x_{q}(t), y_{q}(t)\right)$ denote the x-y UAV coordinate at time $t$. However, for tractability reasons, the flight duration $T$ is divided into $N$ equal time slots, i.e., $T=N \delta_t$, where $\delta_t$ is the duration of each time slot. Hence, the UAV trajectory $\boldsymbol{q}{\left(t\right)}$ during $T$ can be efficiently approximated by a $N$-length sequence $\boldsymbol{q}_{\left[n\right]} = [x_{q[n]}, y_{q[n]}, H_u]$, $n\in \mathcal{N}, \mathcal{N}=\{1,\ldots,N\}$, where $\left(x_{q[n]}, y_{q[n]}\right)$ denote the x-y UAV coordinate at $n$-th time slot. Moreover, to derive the total number of time slots $N$ associated with the UAV trajectory, it is essential to consider both the battery capacity of the available UAV battery and its overall power consumption. Specifically, given a UAV with a battery capacity $B_c$ and an average power consumption per time slot of $\Tilde{P}_d$, the flight duration in time slots can be formulated as
\begin{equation}
	N =\left\lfloor\frac{B_c}{\Tilde{P}_{d}}\right\rfloor,
\end{equation}
where $\lfloor \cdot \rfloor$ is the floor function, and $d \in \{\mathrm{RIS, FD}\}$ describes the communication equipment mounted on the UAV (i.e., RIS or FDR).

\subsection{System's Achievable Rate}

In order to deduce which is the most appropriate communication technology between RISs or FDRs to be mounted on a UAV, a thorough evaluation of the network performance is imperative. Therefore, below, we express the achievable rate of a network when the GN-BS communication is facilitated by either a UAV-mounted RIS or a UAV-mounted FDR. 

\subsubsection{UAV-mounted RIS}

Over the last years, RISs have emerged as efficient tools for manipulating EM waves with minimal energy consumption. Therefore, the integration of RISs on UAVs can offer a dynamic approach to wireless networks, as a UAV can optimally position an RIS to steer incoming EM waves directly to the BS, ensuring flexible 3D network coverage. Hence, considering a reflectarray-based RIS that performs perfect beam-steering \cite{basar2019}, the achievable rate of the $k$-th GN in a UAV-mounted RIS-assisted TDMA network during the $n$-th time slot is expressed as
\begin{equation}
	R_{\mathrm{RIS}, k[n]} = B {a_{k[n]}} \mathrm{log}_2 \left(1 + {\ell}_{\mathrm{RIS}[n]} {\gamma_{t[n]}} G M^2 \right) \! ,
\end{equation}
where $B$ is the system's bandwidth and ${a_{k[n]}} $ is a binary variable that represents the time slot allocation in the TDMA network. Specifically, when ${a_{k[n]}} =1$, it signifies that communication is established between the $k$-th GN and the BS at the $n$-th time slot. Moreover, ${{\ell}_{\mathrm{RIS}[n]}}$ is the path loss that corresponds to the GN-RIS and RIS-BS links, respectively, which can be modeled through the double path loss model as  
\begin{equation}
	{\ell}_{\mathrm{RIS}[n]} =  	{{\ell}_{\mathrm{1}[n]}} 	{{\ell}_{\mathrm{2}[n]}} =\frac{{C_0} {d_0}^{n_p}}{\left({d_{1[n]}}\right)^{n_p}} \times\frac{{C_0}{d_0}^{n_p}}{\left({d_{2[n]}}\right)^{n_p}},
\end{equation}
where $n_p$ is the path-loss exponent, $C_0= \left(\frac{\lambda}{4 \pi}\right)^{2}$ is the path loss of GN-UAV and UAV-BS links at the reference distance $d_0$ with $\lambda$ denoting the wavelength, while ${d_{1[n]}}$ and ${d_{2[n]}}$ express the distances of the GN-UAV and the UAV-BS links at the $n$-th time slot, respectively, and are equal to
\begin{equation}
    d_{1[n]}=\| \boldsymbol{q}_{[n]} - \boldsymbol{l}_{\boldsymbol{k}[n]} \|,
\end{equation}
and
\begin{equation}
    d_{2[n]}=\| \boldsymbol{l}_{\mathrm{\mathbf{BS}}[n]} - \boldsymbol{q}_{[n]} \|,
\end{equation}
with $\| \cdot \|$ being the Euclidean norm. Moreover, $M$ denotes the number of the RIS reflecting elements, $\gamma_{t[n]}=\frac{P_{t[n]}}{\sigma^2}$ is the transmit SNR at the $n$-th time slot, with $P_{t[n]}$ referring to the GN transmit power at the $n$-th time slot and $\sigma^2$ referring to the additive white Gaussian noise (AWGN) affecting the BS, and $G = G_t G_r$ is the product of the GN and BS antenna gains. Finally, considering the LoS nature of air-to-ground communication links, the path loss exponent $n_p$ is equal to $2$. Consequently, the system's achievable rate is equal to
\begin{equation}
	R_{\mathrm{RIS}, k[n]} = B {a_{k[n]}} \mathrm{log}_2 \left(1 + \underbrace{\frac{{\gamma_{t[n]}} G {C_0}^2 {d_0}^{4}  M^2}{\left({d_{1[n]}} {d_{2[n]}}\right)^{2}}}_{{\gamma^{\mathrm{RIS}}_{r,k[n]}} } \right),
\end{equation}
where ${\gamma^{\mathrm{RIS}}_{r,k[n]}}$ is the SNR at the receiver side for the UAV-mounted RIS case when the $k$-th GN is served. It should be mentioned that, in this work, we assume that all $K$ GNs transmit with the same constant low power within the trajectory duration, thus $P_{t[n]}=P_t$ and ${\gamma_{t[n]}}={\gamma_t}$. 

\subsubsection{UAV-mounted FDR}

An alternative solution that has been proposed for air-to-ground networking with improved spectral efficiency is to mount an FDR upon the UAV, referred to as UAV-mounted FDR \cite{bingli2021}. Specifically, a UAV-mounted FDR with $A_n = A_r + A_t$ antennas can provide improved spectral efficiency as it is able to utilize simultaneously $A_r$ antennas for reception and $A_t$ antennas for transmission within one time slot, in contrast to conventional half-duplex relays, which utilize all of their antennas for distinct reception or transmission. However, it is imperative to note that in contrast to RISs, FDRs are inherently susceptible to SI, indicating the importance of advanced SI suppression techniques to optimize their operation. Nevertheless, the process of both analog and digital SI mitigation strategies has positioned FD relays for greater prominence in future network architectures \cite{bingli2021}. To this end, assuming that the UAV-mounted FDR employs the DF protocol and that $A_r$ antennas perform maximum ratio combining (MRC) and $A_t$ antennas perform maximum ratio transmission (MRT) \cite{talebi2008}, the achievable rate of the $k$-th GN within the $n$-th time slot can be expressed as
\begin{equation}
	R_{\mathrm{FD}, k[n]} = \mathrm{min}\left( R_{\mathrm{FD, 1[n]}} , R_{\mathrm{FD, 2[n]}} \right),
\end{equation}
where $R_{\mathrm{FD,1[n]}}$ denotes the achievable rate from the $k$-th GN to the UAV-mounted FDR at the $n$-th time slot, and $R_{\mathrm{FD,2[n]}}$ denotes the achievable rate from the UAV-mounted FDR to the BS at the $n$-th time slot, and can be described as
\begin{equation}
	R_{\mathrm{FD,1[n]}} = B {a_{k[n]}} \log_2 \left(1 + \frac{P_t {\ell}_{\mathrm{1}[n]} G_t A_r}{A_t G_r {P_{u[n]}} \omega + {\sigma_1}^2} \right) ,
\end{equation}
and 
\begin{equation}
	R_{\mathrm{FD,2[n]}} = B {a_{k[n]}} \log_2 \left(1 + \frac{A_t G_r {P_{u[n]}}  {{\ell}_{\mathrm{2}[n]}} }{\sigma_2^2} \right),
\end{equation}
where $P_{u[n]}$ is the FDR transmit power at the $n$-th time slot, $\sigma_1^2=\sigma_2^2=\sigma^2$ denote the variance of the AWGN affecting the FDR and the BS, respectively, and $\omega \in \left[0,1 \right]$ is the self-interference cancellation (SIC) coefficient. Finally, recalling the favorable characteristics of air-to-ground communication links, i.e., $n_p=2$, then $R_{\mathrm{FD,1}[n]}$ and $R_{\mathrm{FD,2}[n]}$ can be rewritten as
\begin{equation}
	R_{\mathrm{FD,1}[n]} \! = \! B {a_{k[n]}} \log_2 \left(\! 1 + \underbrace{\frac{P_t  {C_0} {d_0}^{2} G_t A_r}{{{d_{1[n]}}}^2 \left(A_t G_r P_{u[n]} \omega + {\sigma}^2 \right)}}_{ {\gamma}_{1[n]}} \! \right) \! ,
\end{equation}
and 
\begin{equation}
	R_{\mathrm{FD,2}[n]} = B {a_{k[n]}} \log_2 \left(1 + \underbrace{\frac{A_t G_r P_{u[n]} {C_0} {d_0}^{2} }{{{d_{2[n]}}}^2 {\sigma}^2} }_{ {\gamma}_{2[n]}}\right).
\end{equation}
In addition, the SNR at the receiver side for the UAV-mounted FDR case when the $k$-th GN is served is equal to
\begin{equation}
{\gamma^{\mathrm{FD}}_{r,k[n]}}= \mathrm{min}\left( {\gamma}_{1[n]} , {\gamma}_{2[n]} \right).
\end{equation}
It should be mentioned that similarly with the UAV-mounted RIS case, we assume that the GNs transmission power $P_{t[n]}$ is constant within the UAV flight, i.e., $P_{t[n]}=P_t$.
\begin{remark}
    By setting $ {\gamma}_{1[n]} =  {\gamma}_{2[n]}$ we can derive the optimal FDR transmission power that maximizes the achievable rate of the $k$-th GN at the $n$-th time slot, which is given as
    \begin{equation}
        P^{*}_{u[n]}= \frac{- {d_{1[n]}} \sigma^2 +\sigma \sqrt{{d_{1[n]}}^{2}\sigma^2 + 4 \omega P_t {d_{2[n]}}^{2}G_t A_r} }{2 A_t G_r \omega {d_{1[n]}}}.
    \end{equation}
\end{remark}

\subsection{UAV Power Consumption}\label{energymodel}

Given the inherent battery constraints of UAVs that result in finite flight duration, it becomes crucial to understand their power dynamics, particularly when integrating various communication technologies. Specifically, the UAV power consumption with varied communication equipment can be described as
\begin{equation}
    P_{d[n]}= P_{\mathrm{th},d[n]}+ P_{\mathrm{c},{d}} + P_{\mathrm{tr}},
\end{equation}
where $P_{\mathrm{th},{d}[n]}$ refers to UAV thrusting and encompasses the power demands for transitioning, countering wind drag, and related activities. Furthermore, $P_{\mathrm{c},{d}}$ denotes the power required by the communication equipment (e.g., RIS or FDR), while $P_{\mathrm{tr}}$ is a minimal constant power associated with the UAV's navigational communication (typically less than 1 W), and can be considered negligible \cite{tyrovolas2023}. To elaborate further on the required power for the consumption model, the required power for the reflectarray-based RIS operation $P_{\mathrm{c},\mathrm{RIS}}$ can be described as 
\begin{equation}
    P_{\mathrm{c},\mathrm{RIS}}=MP_{\mathrm{e}}+P_{\mathrm{ct}},
\end{equation}
where $P_{\mathrm{e}}$ represents the power consumption of each reflecting element, and $P_{\mathrm{ct}}$ refers to the energy required by the RIS controller to periodically adjust the RIS phase shift profile within every time slot. Furthermore, the power consumption for the FDR operation $P_{\mathrm{c},\mathrm{FD}}$ is given as
\begin{equation}
    P_{\mathrm{c},\mathrm{FD}}= P_{u^*[n]} \left(1 + \alpha \right) + A_n {P^{C}_{R}},
\end{equation}
where $P_u$ is the FDR transmission power, $\alpha$ is the inverse of the power amplifier drain efficiency, and ${P^{C}_{R}}$ denotes the power consumption of an $A_n$-antenna transceiver \cite{jianhui2022}. This transceiver consumption encompasses the mixer power, the power of phase shifters for each antenna during transmission and reception, the power of each low-noise amplifier per antenna, the frequency synthesizer power, and the encoder power consumption.

Considering the inherent dynamics of UAVs, the thrusting power $P_{\mathrm{th},{d}}$ plays a pivotal role in the total power consumption. Notably, $P_{\mathrm{th},{d}}$ varies within each time slot, being heavily influenced by the UAV speed, its weight, aerodynamic design, and other onboard components such as the battery weight.
Thus, regarding the consumption model presented in \cite{tyrovolas2023}, $P_{\mathrm{th},d[n]}$ can be reliably characterized as
\begin{equation}\label{pthr}	P_{\mathrm{th},d[n]}=C_1 {W_{d[n]}}^2+C_2W_{d[n]}+C_3,
\end{equation}
where $C_1$, $C_2$, and $C_3$ are motor-dependent parameters, while $W_{d[n]}$ encompasses all weight components impacting thrusting power, which can be expressed as
\begin{equation}\label{w}
	W_{d[n]}=U_{\mathrm{w}} + D_{\mathrm{w}} + R_{\mathrm{w},d} + S_{\mathrm{w},d[n]}.
\end{equation}
In more detail, $U_{\mathrm{w}}$ expresses the weight of the UAV frame and its battery, while $D_{\mathrm{w}}$ describes the wind drag given as
\begin{equation}
	D_{\mathrm{w}}=\frac{\rho_a v_a^2 C_d A_{\mathrm{UAV}}}{2g},
\end{equation}
where $\rho_a$ is the air density, $g$ is the gravity acceleration, $v_a$ is the average wind velocity, $C_d$ is the drag shape coefficient given experimentally, and $A_{\mathrm{UAV}}$ is UAV frame area. Moreover, $R_{\mathrm{w},d}$ is the communication equipment weight, where in the RIS case is given by $R_{\mathrm{w},\mathrm{RIS}}=ME_{\mathrm{w}}$, while for an FDR with $A_n$ antennas is equal to $R_{\mathrm{w},\mathrm{DF}}=A_n A_{\mathrm{w}}$, where $E_{\mathrm{w}}$ is the weight of one reflecting element and $A_{\mathrm{w}}$ is the weight of each FDR antenna, respectively. Finally, $S_{\mathrm{w},d[n]}$ is the extra weight added to the motors due to any change in the speed of the UAV in each time slot given by 
\begin{equation}
\label{swd}
	S_{\mathrm{w},d[n]}=(T_{\mathrm{max}}-U_{\mathrm{w}}- D_{\mathrm{w}} -R_{\mathrm{w},d})\frac{\upsilon_{[n]}}{\upsilon_{\mathrm{max}}}, 
\end{equation}
with $T_{\mathrm{max}}$ being the maximum achievable thrust, $\upsilon_{[n]}= \frac{\|\boldsymbol{q}_{[n]}-\boldsymbol{q}_{[n-1]}\|}{\delta_t}$ reflecting the average UAV speed within the $n$-th time slot, and $v_{\mathrm{max}}$ expressing the maximum achievable UAV speed. It should be mentioned that the choice of UAV motors is directly influenced by the weight they are required to lift. As such, a UAV-mounted RIS, which is generally heavier, would demand different motors than a UAV-mounted FDR, to stay in the air for the same amount of time slots as the UAV-mounted FDR. Finally, considering (1) and the available $B_c$, by setting $\upsilon_{[n]}=0$ or $\upsilon_{[n]}=\upsilon_{\mathrm{max}}$ in (20), we can obtain the maximum and the minimum flight duration in terms of time slots, i.e., $N_{\mathrm{max}}$, $N_{\mathrm{min}}$.
\section{Energy-Aware Trajectory Design}\label{section:energy}
In this section, we formulate and solve an optimization problem to derive a UAV trajectory that maximizes the minimum data rate across $K$ GNs, ensuring GN fairness across the network. Unlike existing works, our approach underscores the importance of energy awareness, especially given the inherent battery constraints of UAVs. Specifically, our optimization problem intrinsically incorporates the UAV's power consumption, which is influenced by the mounted communication device, as well as the UAV velocity. Considering these energy dynamics, the formulated optimization problem not only ensures the practical relevance of our findings, but also provides design insights into the characteristics of the employed RIS and FDR that optimize network performance. Moreover, the optimization guarantees that the power consumed by the UAV during its trajectory does not surpass its available battery energy, ensuring the UAV's safe return to the BS and highlighting the significance of energy-aware trajectory design. To this end, we will examine the network efficiency in terms of achievable rate and fairness for both RIS and FDR.
\vspace{-5mm}
\subsection{Problem Formulation}
To efficiently maximize the minimum data rate of the network, we aim to jointly optimize the UAV trajectory and TDMA user scheduling, taking into account mobility, user scheduling, and UAV power consumption constraints. Given these considerations, by leveraging the integer variable $\boldsymbol{A}$ that represents the TDMA scheduling, and the continuous variable $\boldsymbol{Q}$ that describes the UAV trajectory, the optimization problem for both UAV-mounted RIS and UAV-mounted FDR cases can be formulated as
\begin{equation} \tag{\textbf{P1}}\label{opt1}
\begin{aligned}
        &\underset{{\boldsymbol{A,Q}}}{\textbf{max}}\,\underset{k}{\min} \left\{ \sum_{n=1}^N R_{d, k[n]}\right\}\\
        \textbf{s.t} \,\,\,\, &\mathrm{C}_1: \,\boldsymbol{q}_{[1]} = \boldsymbol{q}_{[N]}, \\
        &\mathrm{C}_2: \, \upsilon_{[n]} \leq \upsilon_{\mathrm{max}}, \ \forall n \in \mathcal{N},\\
        &\mathrm{C}_3: \, \sum _{k=1} ^ K a_{k[n]}  \leq 1, \ \forall n \in \mathcal{N},\\
        &\mathrm{C}_4: \, B_c - \delta_t \sum _{n=1} ^N P_{d[n]} \geq 0, \\
        &\mathrm{C}_5: \,a_{k[n]}  \in \{0, 1\}, \ \forall n \in \mathcal{N},\\
        &\mathrm{C}_6: \, q_\mathrm{min} \leq \boldsymbol{q}_{[n]} \leq q_\mathrm{max}, \ \forall n \in \mathcal{N},\\
        &\mathrm{C}_7: \, P_{u[n]} \leq P_\mathrm{max}, \ \forall n \in \mathcal{N},
\end{aligned}
\end{equation}
where  $R_{d, k[n]}$ is the GN data rate which is given by
\begin{equation} \label{OPBTS}
\small
\begin{split}
R_{d, k[n]} =
    \begin{cases}
     0, & {\gamma^{d}_{r,k[n]}}  <  \gamma_{\mathrm{thr}} \\
    B {a_{k[n]}} \mathrm{log}_2 \left(1 + \gamma^{\mathrm{RIS}}_{r,k[n]} \right),   &   {\gamma^{d}_{r,k[n]}}  > \gamma_{\mathrm{thr}} , d=\text{RIS} \\
         B {a_{k[n]}} \mathrm{log}_2 \left(1 + \gamma^{\mathrm{FD}}_{r,k[n]} \right),   &  {\gamma^{d}_{r,k[n]}}  > \gamma_{\mathrm{thr}} , d=\text{FDR} ,
    \end{cases}
\end{split}
\end{equation}
 where $\gamma_{\mathrm{thr}}$ is an SNR threshold. In more detail, $\mathrm{C}_1$ forces the UAV trajectory to begin and end at the same point. In addition, $\mathrm{C}_2$ indicates that the  UAV speed $\upsilon_{[n]}$ cannot exceed the maximum UAV velocity $\upsilon_{\mathrm{max}}$, while $\mathrm{C}_3$ defines that only one GN-BS pair can be served by the UAV-mounted RIS or the UAV-mounted FDR within a certain time slot. Moreover, $\mathrm{C}_4$ indicates that the UAV's power consumption during its trajectory must not exceed its battery's available energy, while $\mathrm{C}_5$ and $\mathrm{C}_6$ set the lower and upper bounds for the optimization variables $\boldsymbol{a,q}$, with $\mathrm{C}_6$ ensuring the UAV remains within the predefined rectangular field. Lastly, $\mathrm{C}_7$ describes that the UAV's transmission power should be less than $P_{\mathrm{max}}$, denoting the peak power limit of the UAV, where for the UAV-mounted RIS case, $P_{u[n]}=0$ as the RIS does not consume any transmission power.
 
\subsection{Problem Solution}
As it can be seen, problem (\ref{opt1}) is intractable since it contains both continuous and integer variables while its objective function is non-convex. To this end, given that current optimization methods struggle with problems containing non-linear constraints paired with integer variables, a separation of the integer variable $\boldsymbol{A}$ and the continuous variable $\boldsymbol{Q}$ becomes essential. To address this, the alternate optimization technique is employed, which relies on successively optimizing each optimization variable block until convergence \cite{bingli2021}. Therefore, for a fixed trajectory $\boldsymbol{Q}$ we have
\begin{equation} \tag{\textbf{PA.1}}\label{optA.1}
\begin{aligned}
        &\underset{{\boldsymbol{A}}}{\textbf{max}}\,\underset{k}{\min} \left\{ \sum_{n=1}^N R_{d, k[n]}\right\}\\
        \textbf{s.t} \,\,\,\, &\mathrm{C}_1: \, \sum _{k=1} ^ K a_{k[n]}  \leq 1, \ \forall n \in \mathcal{N},\\
        &\mathrm{C}_2: \,a_{k[n]}  \in \{0, 1\},\ \forall n \in \mathcal{N}.
\end{aligned}
\end{equation}
As it can be seen, problem \eqref{optA.1} is an integer programming problem, however, it is not in canonical form, since the objective function is a non-linear function. Thus, by utilizing the auxiliary variable $r_\mathrm{min}$, problem \eqref{optA.1} is equivalently written as 
\begin{equation} \tag{\textbf{PA.2}}\label{optA.2}
\begin{aligned}
        &\underset{{\boldsymbol{A},r_\mathrm{min}}}{\textbf{max}}\,\,\,\, r_\mathrm{min}\\
        \textbf{s.t} \,\,\,\, &\mathrm{C}_{1}: \, \sum _{k=1} ^ K a_{k[n]}  \leq 1, \ \forall n \in \mathcal{N},\\
        &\mathrm{C}_2: \,a_{k[n]}  \in \{0, 1\},\ \forall n \in \mathcal{N},\\
        &\mathrm{C}_3: \, \sum_{n=1}^N a_{k[n]}R_{d, k[n]} \geq r_\mathrm{min}, \ \forall k \in \mathcal{K},
\end{aligned}
\end{equation}
which is a mixed-integer linear programming problem (MILP), since for given trajectory $\boldsymbol{Q}$, $R_{d, k[n]}$ is constant for either the FDR or the RIS case. As a consequence, \eqref{optA.2} can be optimally solved using off-the-self optimization tools such as the Branch and Bound method.
\subsubsection{Optimization of $\boldsymbol{Q}$} Given the TDMA schedule $\boldsymbol{A}$, problem \eqref{opt1} can be written as 
\begin{equation} \tag{\textbf{PQ.1}}\label{optQ.1}
\begin{aligned}
        &\underset{{\boldsymbol{Q}}}{\textbf{max}}\,\underset{k}{\min} \left\{ \sum_{n=1}^N a_{k[n]} R_{d, k[n]}\right\}\\
        \textbf{s.t} \,\,\,\, &\mathrm{C}_1: \,\boldsymbol{q}_{[1]} = \boldsymbol{q}_{[N]}, \\
        &\mathrm{C}_2: \, \upsilon_{[n]} \leq \upsilon_{\mathrm{max}}, \ \forall n \in \mathcal{N},\\
        &\mathrm{C}_3: \, B_c - \delta _t \sum _{n=1} ^N P_{d[n]}  \geq 0, \\
        &\mathrm{C}_4: \, q_\mathrm{min} \leq \boldsymbol{q}_{[n]} \leq q_\mathrm{max},\ \forall n \in \mathcal{N}, \\
        &\mathrm{C}_5: \, P_{u[n]} \leq P_\mathrm{max}, \ \forall n \in \mathcal{N},
\end{aligned}
\end{equation}
which is a non-convex problem due to the non-concave and dual branch objective function $R_{d,k[n]}$. Furthermore, in \eqref{swd}, $P_{d[n]}$ is influenced by the UAV's speed $\upsilon_{[n]}$, which is derived from the differences in consecutive positions $x_{q[n]}$ and $ y_{q[n]}$, thus, given that these positions are constants for each time slot, the relationship of  $P_{d[n]}$ with $x_q$ and $y_q$ is affine, as it comprises linear differences and constant terms. Thus, $P_{d[n]}$ is by definition convex which makes constraint $\mathrm{C}_3$ convex as well. In addition, by introducing the auxiliary variable $r_\mathrm{min}$, problem \eqref{optQ.1} is equivalently transformed as follows
\begin{equation} \tag{\textbf{PQ.2}}\label{optQ.2}
\begin{aligned}
        &\underset{\boldsymbol{x}_q, \boldsymbol{y}_q,r_\mathrm{min}}{\textbf{max}}\,\,\, r_\mathrm{min}\\
        \textbf{s.t} \,\,\,\, &\eqref{optQ.1}: \mathrm{C}_1,\mathrm{C}_2,\mathrm{C}_3,\mathrm{C}_4, \mathrm{C}_5\\
        &\mathrm{C}_6: \,\sum_{\substack{n=1,\\a_{k[n]}=1}}^N\!\!R_{d, k[n]} \geq r_\mathrm{min}, \ \forall k \in \mathcal{K}.
\end{aligned}
\end{equation}
Again, problem \eqref{optQ.2} is still non-convex due to $\mathrm{C}_6$. To tackle this, we can convert it from a dual branch function into a single function by introducing an appropriate constraint that assures that the UAV always serves a GN for which the received SNR at the BS-side is above $\gamma_{\mathrm{thr}}$. Additionally, $\forall n \in \mathcal{N}$ and  $\forall k \in \mathcal{K}$ for which $a_{k[n]}=1$, we can introduce the auxiliary variables $r_{k[n]}$, thus the problem \eqref{optQ.2} can be rewritten as 
\begin{equation} \tag{\textbf{PQ.3}}\label{optQ.3}
\begin{aligned}
        &\underset{\boldsymbol{x}_q, \boldsymbol{y}_q,r_\mathrm{min}, r_{k[n]}}{\textbf{max}}\,\,\, r_\mathrm{min}\\
        \textbf{s.t} \,\,\,\, &\eqref{optQ.1}: \mathrm{C}_1,\mathrm{C}_2,\mathrm{C}_3,\mathrm{C}_4,\mathrm{C}_5\\
        &\mathrm{C}_6: \,\sum_{\substack{n=1,\\a_{k[n]}=1}}^N\!\! r_{k[n]} \geq r_\mathrm{min}, \ \forall k \in \mathcal{K}\\
        &\mathrm{C}_7: \, {\gamma^{d}_{r,k[n]}} \geq {\gamma_{\mathrm{thr}}}, \ \forall n \in \mathcal{N},  \ \forall k \in \mathcal{K}\,\,\text{that}\,\, a_{k[n]}=1, \\
        &\mathrm{C}_8: \, R_{d, k[n]} \geq r_{k[n]}, \ \forall n \in \mathcal{N}, \ \forall k \in \mathcal{K}\,\,\text{that}\,\, a_{k[n]}=1. 
\end{aligned}
\end{equation}
Due to the distinct achievable rates of the UAV-mounted RIS and the UAV-mounted FDR, $\mathrm{C}_7$ and $\mathrm{C}_8$ have to be dealt differently for each case. To this end, the distinct approaches for the two cases are presented below.
\begin{itemize}
    \item \textbf{UAV-mounted RIS}: Utilizing (6) and defining $A_\mathrm{R} = {\gamma_t} G {C_0}^2 {d_0}^{4}  M^2$, we adopt the approximation $\log_2(1+z) \approx \log_2(z)$. This choice is motivated by the dominant high SNR conditions inherent in our system, particularly under LoS scenarios. In fact, given the UAV's strategic positioning relative to the GNs it serves, $z$ is typically large, making the approximation increasingly pertinent. Thus, $\mathrm{C}_8$ can be expressed as
    \begin{equation}
    \begin{aligned}
        &\mathrm{C}_8: \log_2({d_{1[n]}}^2) + \log_2\left(\frac{{d_{2[n]}}^2}{A_\mathrm{R}}\right) \leq -r_{k[n]}.
    \end{aligned}
    \end{equation}
Furthermore, by introducing the auxiliary variables $s_{k,n}$ and $t_{k[n]}$, such that $\log_2({d_{1[n]}}^2)\leq s_{k,n}$, and $\log_2({d_{2[n]}}^2)\leq t_{k[n]}$, $\mathrm{C}_6$ and $\mathrm{C}_8$ are rewritten as 
    \begin{equation}
        \begin{aligned}
        &\mathrm{C}_6: \sum_{n=1}^N \left( s_{k[n]} + t_{k[n]} \right) \leq -r_\mathrm{min},\\
        &\mathrm{C}_8: s_{k[n]} + t_{k[n]} \leq -r_{k[n]},
        \end{aligned}
    \end{equation}
    while the following constraints occur as well:
        \begin{equation}
        \begin{aligned}
        &\mathrm{C}_{8.A}:(x_{q[n]}-x_k)^2 + (y_{q[n]}-y_k)^2 + H_u^2 - 2^{s_{k,n}} \leq 0 \\
        &\mathrm{C}_{8.B}: \frac{(x_{q[n]}-x_b)^2 + (y_{q[n]}-y_b)^2 + (H_u-H_{\mathrm{BS}})^2}{A_\mathrm{R}} \\
        &\quad\quad\quad- 2^{t_{k[n]}} \leq 0.
        \end{aligned}
    \end{equation}
We note that constraint $\mathrm{C}_7$ can be handled in the same way, by introducing the auxiliary variables $\hat{s}_{k[n]}$ and $\hat{t}_{k[n]}$. Then, the convex form of $\mathrm{C}_7$ is equivalently given as follows
\begin{equation}
    \begin{aligned}
        &\mathrm{C}_{7}: \hat{s}_{k[n]} + \hat{t}_{k[n]} \leq -\log(\gamma_{\mathrm{thr}}),\\
        &\mathrm{C}_{7.A}:(x_{q[n]}-x_k)^2 + (y_{q[n]}-y_k)^2 + H_u^2 - 2^{\hat{s}_{k[n]}} \leq 0 \\
        &\mathrm{C}_{7.B}: \frac{(x_{q[n]}-x_b)^2 + (y_{q[n]}-y_b)^2 + (H_u-H_{\mathrm{BS}})^2}{A_\mathrm{R}} \\
        &\quad\quad\quad- 2^{\hat{t}_{k[n]}} \leq 0.
    \end{aligned}
\end{equation}
To address the non-convexity introduced by the new constraints \(\mathrm{C}_{7.A}\), \(\mathrm{C}_{7.B}\), \(\mathrm{C}_{8.A}\), \(\mathrm{C}_{8.B}\), the successive approximation method (SCA) is employed \cite{bingli2021}. Thus, by substituting the non-convex terms with their first-order Taylor approximation, we formulate the convex optimization problem for the trajectory design in the UAV-mounted RIS scenario as follows:
   \begin{equation} \tag{\textbf{PQ.4-RIS}}\label{optQ.4}
\begin{aligned}
        &\underset{\substack{\boldsymbol{x}_q, \boldsymbol{y}_q,r_\mathrm{min},\\ r_{k[n]}, s_{k,n},t_{k[n]}}}{\textbf{max}}\,\,\, r_\mathrm{min}\\
        \textbf{s.t} \,\,\,\, &\eqref{optQ.3}: \mathrm{C}_1,\mathrm{C}_2,\mathrm{C}_3,\mathrm{C}_4,\mathrm{C}_5, \mathrm{C}_6, \mathrm{C}_7, \mathrm{C}_8\\
        &\mathrm{C}_{7.A}:(x_{q[n]}-x_k)^2 + (y_{q[n]}-y_k)^2 + {H_u}^2 \\
        & \quad\quad - 2^{\hat{s}_{k[n],0}} - (\hat{s}_{k[n]}-\hat{s}_{k[n],0})2^{\hat{s}_{k[n],0}} \leq 0 \\
        &\mathrm{C}_{{7.B}}: \frac{(x_{q[n]} \!-\!x_b)^2 + (y_{q[n]} \!-\! y_b)^2 + (H_u \!-\! H_{\mathrm{BS}})^2}{A_\mathrm{R}} \\
        & \quad\quad - 2^{\hat{t}_{k[n],0}} - (\hat{t}_{k[n]}-\hat{t}_{k[n],0})2^{\hat{t}_{k[n],0}} \leq 0,\\
        &\mathrm{C}_{8.A}:(x_{q[n]}-x_k)^2 + (y_{q[n]}-y_k)^2 + {H_u}^2 \\
        & \quad\quad - 2^{s_{k[n],0}} - (s_{k,n}-s_{k[n],0})2^{s_{k[n],0}} \leq 0 \\
        &\mathrm{C}_{{8.B}}: \frac{(x_{q[n]} \!-\!x_b)^2 + (y_{q[n]} \!-\! y_b)^2 + (H_u \!-\! H_{\mathrm{BS}})^2}{A_\mathrm{R}} \\
        & \quad\quad - 2^{t_{k[n],0}} - (t_{k[n]}-t_{k[n],0})2^{t_{k[n],0}} \leq 0,
\end{aligned}
\end{equation}
where $s_{k[n],0}$, $t_{k[n],0}$, $\hat{s}_{k[n],0}$, $\hat{t}_{k[n],0}$ are the arbitrary initial points for the Taylor approximation. 
\begin{remark}
For the case in the UAV-mounted RIS scenario where the trajectory's initial point coincides with the BS location and the path loss exponents of the BS-RIS and RIS-GN channels are equal, the optimal trajectory is to hover at this point until the UAV's battery is depleted. This strategy maximizes the network's data rate by achieving the minimum overall path loss through minimal UAV-BS distance due to the double path loss phenomenon, and simultaneously minimizes energy consumption, capitalizing on the UAV's hovering state.
\end{remark}
\color{black}
    \item \textbf{UAV-mounted FDR}: For the UAV-mounted FDR case, the constraint $\mathrm{C}_7$ can be equivalently written as 
\begin{equation}
\begin{aligned}
    \mathrm{C}_{7.A}:&\frac{(x_{q[n]}-x_k)^2 + (y_{q[n]}-y_k)^2 + {H_u}^2}{A_{\mathrm{FD},1[n]}} \leq \frac{1}{\gamma_\mathrm{thr}}\\
    \mathrm{C}_{7.B}:& \frac{(x_{q[n]}-x_b)^2 + (y_{q[n]}-y_b)^2 + (H_u-H_{\mathrm{BS}})^2}{A_{\mathrm{FD},2[n]}} \\ &\leq \frac{1}{\gamma_\mathrm{thr}},
\end{aligned}
\end{equation}
which is convex. Furthermore, by utilizing (7), (10), and (11), and considering that the condition $\mathrm{min}\{x,y\}\geq t$ implies that both $x\geq t$ and $y\geq t$, thus, $\mathrm{C}_8$ in $\textbf{PQ.3}$ can be equivalently divided into two separate constraints:
    \begin{equation}
        \begin{aligned}
            &\mathrm{C}_{8.A}: \log_2\left(\mathrm{A}_{\mathrm{FD},1[n]}{d_{1[n]}}^{-2}\right) \geq  r_{k[n]},\\
            &\mathrm{C}_{8.B}: \log_2\left(\mathrm{A}_{\mathrm{FD},2[n]}{d_{2[n]}}^{-2}\right) \geq  r_{k[n]},
        \end{aligned}
    \end{equation}
where $\mathrm{A}_{\mathrm{FD},1[n]} = \frac{P_t  {C_0} {d_0}^{2} G_t A_r}{A_t G_r P_{u[n]} \omega + {\sigma}^2 } $, and $\mathrm{A}_{\mathrm{FD},2[n]} = \frac{A_t G_r P_{u[n]} {C_0} {d_0}^{2}}{ \sigma^2}$. Finally, similarly to the UAV-mounted RIS case, we utilize the SCA method, thus the trajectory optimization problem for the UAV-mounted FDR case is given as
\begin{equation} \tag{\textbf{PQ.4-FDR}}\label{optQ.fdr}
    \begin{aligned}
        &\underset{\boldsymbol{x}_q, \boldsymbol{y}_q, r_\mathrm{min}, r_{k[n]}}{\textbf{max}}\,\,\, r_\mathrm{min}\\
        \textbf{s.t} \,\,\,\, &\eqref{optQ.3}: \mathrm{C}_1,\mathrm{C}_2,\mathrm{C}_3,\mathrm{C}_4,\mathrm{C}_5,\mathrm{C}_6, \mathrm{C}_{7.A}, \mathrm{C}_{7.B}\\
        &\mathrm{C}_{8.A}:\frac{(x_{q[n]}-x_k)^2 + (y_{q[n]}-y_k)^2 + {H_u}^2}{A_{\mathrm{FD},1[n]}} \\
        & \quad - 2^{r_{k[n],0}} - (r_{k[n]}-r_{k[n],0})2^{r_{k[n],0}} \leq 0 \\
        &\mathrm{C}_{8.B}: \frac{(x_{q[n]} \!-\! x_b)^2 + (y_{q[n]} \!-\! y_b)^2 + (H_u \!-\! H_{\mathrm{BS}})^2}{A_{\mathrm{FD},2[n]}} \\
        & \quad - 2^{r_{k[n],0}} - (r_{k[n]}-r_{k[n],0})2^{r_{k[n],0}} \leq 0.
    \end{aligned}
\end{equation}
\end{itemize}
As it can be observed, both problem \eqref{optQ.4} and problem \eqref{optQ.fdr} are now convex, allowing them to be addressed using standard optimization techniques like the interior-point method. The procedure for the joint TDMA-trajectory design, for both the UAV-mounted RIS and UAV-mounted FDR cases, is outlined in Algorithm 1. It is worth noting that the values for $\mathrm{iter}_1$ and $\mathrm{iter}_2$ are selected to ensure that the solutions from both the SCA and alternate optimization methods converge to a consistent solution \cite{bingli2021}, which is then presented as the final output of Algorithm 1. 
\begin{algorithm}
\linespread{1}\selectfont
\begin{algorithmic}[1]\label{alg1}
\caption{Trajectory Design for UAV-mounted RIS or UAV-mounted FDR} 
\State {Initialize $\mathrm{iter}_1$, $\mathrm{iter}_2$ $B_c$, $\delta_t$, $v_{\mathrm{max}}$, $U_{\mathrm{w}}$, $D_{\mathrm{w}}$, $R_{\mathrm{w},d}$ }
\For{ {$N=N_{\mathrm{min}}, N_{\mathrm{min}}+1,....,N_{\mathrm{max}}$}} 
   \State {Initialize $\boldsymbol{A}_\mathrm{init}$, $\boldsymbol{Q}_\mathrm{init}$}
    \For{ {$i=0, 1, 2,..., \mathrm{iter}_1$}} 
        \State{For $\boldsymbol{Q}_\mathrm{init}$, solve \eqref{optA.2} and obtain $\boldsymbol{A}^{i}$} 
        \For{ {$j=0, 1, 2,..., \mathrm{iter}_2$}}
            \State{For $\boldsymbol{A}^{i}$, solve  \eqref{optQ.fdr} or \eqref{optQ.4} \hspace{5cm} \hspace*{1cm} and obtain $\boldsymbol{Q}^{j}$}
            \State{$r_{k[n],0} \gets r^{j}_{k[n]}$, $s_{k[n],0} \gets s^{j}_{k[n]}$, $t_{k[n],0} \gets t^{j}_{k[n]}$}
        \EndFor
        \State{$\boldsymbol{Q}_\mathrm{init} \gets \boldsymbol{Q}^{\mathrm{iter}_2}$}
    \EndFor
\State{$\boldsymbol{Q}^\mathrm{*} \gets \boldsymbol{Q}^{\mathrm{iter}_2}$}
\State{For $\boldsymbol{Q}^\mathrm{*}$, solve \eqref{optA.2} and obtain $\boldsymbol{A}^{*}$}
\EndFor
\State{Obtain the best $\boldsymbol{Q}^\mathrm{*}$ and $\boldsymbol{A}^\mathrm{*}$}
\end{algorithmic}
\end{algorithm}

\section{Simulation Results}\label{section:simulation}
\begin{table}\label{Table:1}
	\renewcommand{\arraystretch}{1.00}
	\caption{\textsc{Power Consumption Model Parameters}}
	\label{values}
	\centering
	\begin{tabular}{lll}
		\hline
		\bfseries Parameter & \bfseries Notation & \bfseries Value \\
		\hline\hline
		UAV weight				  &  $U_{\mathrm{w}}$		& $3.25$ kg									\\
		Reflecting element weight		  &  $E_{\mathrm{w}}$		    & $ 3.43 \times 10^{-3}$ kg			        \\
  	Antenna weight		  &  $A_{\mathrm{w}}$		    
        & $ 8 \times 10^{-3}$ kg			        \\
		Battery weight			  &  $B_{\mathrm{w}}$			& $1.35$ kg									\\
		Battery capacity          &  $B_c$  	    & $45$ Wh 		    						\\		
             Reflecting element consumption & $P_n$  & $2$ $\mu$W                                     \\
             Controller consumption & $P_c$  & $50$ mW                                    \\
		Inverse of pow. ampl. drain eff.   &  $\alpha$    & $1.875$ 		    			\\
        Transceiver consumption   &  $P^{C}_{R}$    & $1.5$ W 		    			\\
		Maximum achievable thrust &  $T_{\mathrm{max}}$ 		& $17$ kg		    			\\
		Maximum UAV speed		  &  $\upsilon_{\mathrm{max}}$		& $62$ km/h		  	  				\\		
		Air density				  &  $\rho_a$			& $1.225$ kg/m$^3$  						\\
		Air velocity			  &  $v_a$			& $2.5$ m/s (Light Air)  					\\
		Drag shape coefficient	  &  $C_d$    		&$0.005$	\\
		UAV frame				  &  $A_{\mathrm{UAV}}$		& $0.5 \times 0.5 $ m$^2$ 	\\
  	Reflecting element area				  &         
        $A_{\mathrm{re}}$		& $\frac{\lambda}{10} \times \frac{\lambda}{10} $ m$^2$ 	\\
		Gravity acceleration	  &  $g$			& 9.8 m/s$^2$								\\
        MN505-s KV320 T-MOTOR	  &  $C_1, C_2, C_3$			& 4, 86, -21.2								\\
        AT4130 KV230 T-MOTOR	  &  $C_1, C_2, C_3$			& 10.5, -46, 744								\\
		\hline
	\end{tabular}
\end{table}

In this section, we present numerical results to assess the performance of the proposed UAV-assisted communication scenarios, whose power consumption and network parameters are set as detailed in Table I and Table II, respectively. Specifically, we consider an uplink communication system assisted by i) a UAV-mounted RIS or ii) a UAV-mounted FDR with 10 GNs that transmit with power equal to $P_t=0$ dBm, that are also randomly distributed over a rectangular area with sides equal to $L=750$ m,  unless otherwise stated. In addition, a single-antenna BS is located in the origin of the rectangular area which also serves as a UAV CS. It should be mentioned that, in alignment with practical scenarios, we assume that the UAV starts and concludes its trajectory at the same location, reflecting the common practice, where the UAV takes off and lands at the same place for recharging purposes. Moreover, it is assumed that the UAV-mounted RIS is equipped with the MN505-s KV320 T-MOTOR motors due to its weight, while the lighter UAV-mounted FDR uses the AT4130 KV230 T-MOTOR motors. Moreover, the size of the UAV frame, combined with the size of each reflecting element as outlined in Table I, allows for a maximum of 1600 reflecting elements to be fitted on the UAV frame. In contrast, for the UAV-mounted FDR, the frame can accommodate a maximum of 12 antennas, arranged as a uniform linear array (ULA) with an inter-distance of $\frac{\lambda}{2}$ along the frame's diagonal, for which we assume that $A_t=A_r$. In addition, we consider a benchmark TDMA scheduling scheme where each GN is allocated $N_{\mathrm{GN}}= \left\lfloor\frac{N}{K}\right\rfloor$ time slots, to clearly show the efficiency of our algorithm. To be more precise, in the benchmark scheme, the UAV serves the nearest GN for $N_{\mathrm{GN}}$ time slots and then serves the next nearest unserved GN, continuing until every GN has been served before returning to its initial location. Additionally, the benchmark trajectories are determined through a detailed search for the most effective combination of time slots and trajectory sizes, all considered within the operational limits of the UAV’s battery life. Finally, all of the results were calculated through Monte Carlo simulations with 1000 iterations.

\begin{table}
	\renewcommand{\arraystretch}{1.00}
	\caption{\textsc{Network Parameters}}
	\label{values_sim}
	\centering
	\begin{tabular}{lll}
		\hline
		\bfseries Parameter & \bfseries Notation & \bfseries Value \\
		\hline\hline
		UAV height          &  $H_u$  	    & $100$ m 		    						\\
        BS height          &  $H_{\mathrm{BS}}$  	    & $15$ m 		    						\\
		Number of GNs			  &  $K$			& $10$ 									\\
  	Max transmit FDR power		  &  $P_{\mathrm{max}}$			& $0$ dBm 									\\
        time slot duration		  &  $\delta_t$		& 
        $1$ s		  	  						\\
		Reference distance				  &  $d_0$		& $1$ m									\\
		Bandwidth				  &  $B$		& $1$ MHz 						\\
		Transmit power   &  $P_{t}$    & $0$ dBm \\
        Noise Power   &  $\sigma^2$    & $-144$ dBm 		    						\\
		Antenna gains &  $G_t,G_r$ 		& $0$ dB 		    						\\
		Wavelength		  &  $\lambda$		& $0.125$ m		  	  						\\
  	SIC coefficient		  &  $\omega$		& 
        $-90$ dB		  	  						\\
        Iterations		  &  $\mathrm{iter}_1, \mathrm{iter}_2$		& $20$, $20$		  	  						\\
		\hline
	\end{tabular}
\end{table}

\begin{figure*}
    \centering
    \begin{subfigure}{.32\textwidth}
    {	\begin{tikzpicture}
	\begin{axis}[
	height=0.97\linewidth,
	width=0.97\linewidth,
	xlabel = {Length (m)},
	ylabel = {Width (m)},
	xmin = -375,xmax = 375,
	ymin = -375,
	ymax = 375,
	ytick = {-375,-250,...,375},
	xtick = {-375,-250,...,375},
	grid = major,
	legend image post style={xscale=0.9},
	legend cell align = {left},
      legend style={at={(1,0)},anchor=south east,font = \scriptsize}
	]
	\addplot[
	blue,
        only marks,
	mark=triangle*,
	mark repeat = 1,
	mark size = 3,
	]
	table {Final_dats/fig1/aggregators_position.dat};
  	\addlegendentry{GN}
 	\addplot[
	black,
        only marks,
	mark=*,
	mark repeat = 1,
	mark size = 2,
	]
	table {Final_dats/fig1/BS.dat};
 	\addlegendentry{BS}
	\addplot[
	black,
        no marks,
	line width = 0.75pt,
	style = solid,
	]
	table {Final_dats/fig1/circle_traj.dat};
	\end{axis}
	\end{tikzpicture}}
    \caption{} \label{fig:Circle} 
    \end{subfigure}
    \centering
    \begin{subfigure}{.32\textwidth}{\begin{tikzpicture}
	\begin{axis}[
	height=0.97\linewidth,
	width=0.97\linewidth,
	xlabel = {Length (m)},
	ylabel = {Width (m)},
	xmin = -375,xmax = 375,
	ymin = -375,
	ymax = 375,
	ytick = {-375,-250,...,375},
	xtick = {-375,-250,...,375},
	grid = major,
	legend image post style={xscale=0.9},
	legend cell align = {left},
      legend style={at={(1,0)},anchor=south east,font = \scriptsize}
	]
	\addplot[
	blue,
        only marks,
	mark=triangle*,
	mark repeat = 1,
	mark size = 3,
	]
	table {Final_dats/fig1/aggregators_position.dat};
  	\addlegendentry{GN}
 	\addplot[
	black,
        only marks,
	mark=*,
	mark repeat = 1,
	mark size = 3,
	]
	table {Final_dats/fig1/BS.dat};
 	\addlegendentry{BS}
	\addplot[
	black,
       no marks,
	line width = 0.75pt,
	style = solid,
	]
	table {Final_dats/fig1/rhombus_traj.dat};
	\end{axis}
	\end{tikzpicture}}
    \caption{} \label{fig:Rombus} 
    \end{subfigure}
        \begin{subfigure}{.32\textwidth}{\begin{tikzpicture}
	\begin{axis}[
	height=0.97\linewidth,
	width=0.97\linewidth,
	xlabel = {Length (m)},
	ylabel = {Width (m)},
	xmin = -375,xmax = 375,
	ymin = -375,
	ymax = 375,
	ytick = {-375,-250,...,375},
	xtick = {-375,-250,...,375},
	grid = major,
	legend image post style={xscale=0.9},
	legend cell align = {left},
      legend style={at={(1,0)},anchor=south east,font = \scriptsize}
	]
	\addplot[
	blue,
        only marks,
	mark=triangle*,
	mark repeat = 1,
	mark size = 3,
	]
	table {Final_dats/fig1/aggregators_position.dat};
  	\addlegendentry{GN}
 	\addplot[
	black,
        only marks,
	mark=*,
	mark repeat = 1,
	mark size = 2,
	]
	table {Final_dats/fig1/BS.dat};
 	\addlegendentry{BS}
	\addplot[
	black,
       no marks,
	line width = 0.75pt,
	style = solid,
	]
	table {Final_dats/fig1/spiral_traj.dat};
	\end{axis}
	\end{tikzpicture}}
    \caption{} \label{fig:SPIRAL} 
    \end{subfigure}
    \caption{ Benchmark UAV trajectories: (a) Circle (b) Rombus (c) Spiral.}
    \vspace{-5mm}.
\end{figure*}
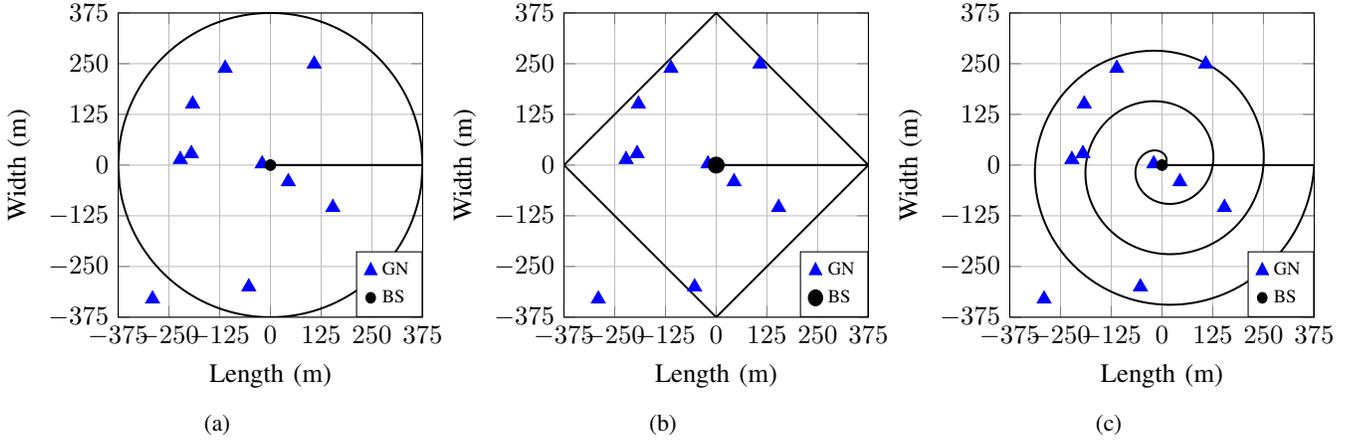

In Fig. 2, three distinct benchmark UAV trajectories are illustrated, each serving as a feasible initial trajectory, denoted as $\boldsymbol{Q}_\mathrm{init}$, for the implementation of Algorithm 1. These trajectories include i) a straight line connecting the origin to the midpoint on the right side of the rectangular field, followed by a circle with radius $\frac{L}{2}$ centered at the axis origin; ii) a straight line connecting the origin to the midpoint on the right side of the rectangle, succeeded by a rhombus with sides measuring $\frac{L\sqrt{2}}{2}$; and iii) an Archimedean spiral described in polar coordinates by $r = \frac{L}{12 \pi} \theta$, with $\theta \in [0, 2 \pi)$, followed by a straight line leading back to the origin. In the circular and rhombus trajectories, the UAV initially follows the straight line, completes the circle or rhombus, and then retraces its path along the straight line back to the origin, while in the spiral trajectory the UAV navigates the spiral path before returning to the origin via the straight line. Notably, in all these trajectories, the UAV navigates counter-clockwise, with the starting and ending points, $\boldsymbol{q}_{[1]}$ and $\boldsymbol{q}_{[N]}$, aligning with the BS location. It should be highlighted that as shown in (1), the calculation of $N$ based on $B_c$ and $P_d$ informs the minimum average speed required for the UAV to deplete its battery at the last point of the trajectory, thereby defining the feasible size of each trajectory within the allocated time slots. Specifically, the maximum number of time slots corresponds to the case where the UAV minimizes its power consumption (i.e., hovering), while reducing the number of time slots allows for larger trajectories. This indicates that an optimal number of time slots must be determined, as too many slots could restrict the UAV from performing large enough trajectories due to energy limitations, while fewer slots may limit service duration. Thus, the evaluation of different initial trajectories, each corresponding to a specific number of time slots is demanded, which can be done by initializing different values of $N$ in step 1 of Algorithm 1 and selecting the value of $N$ maximizing $r_\mathrm{min}$.

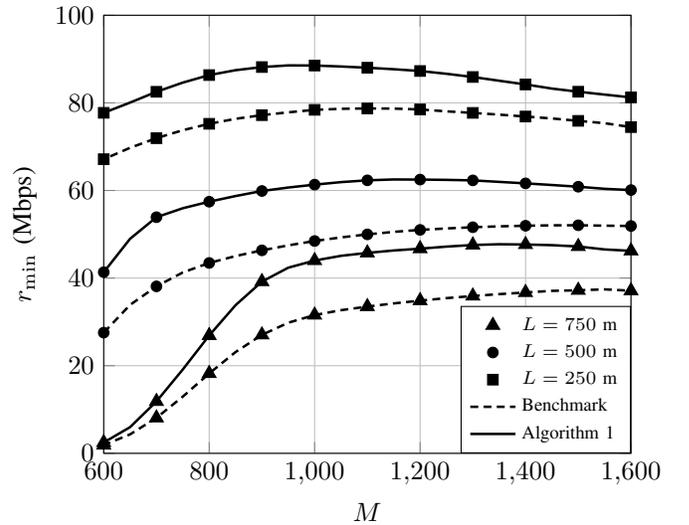
\begin{figure}
        \centering
        \begin{tikzpicture}
	\begin{axis}[
	width=0.97\linewidth,
	xlabel = {$M$},
	ylabel = {$r_\mathrm{min}$ (Mbps)},
	xmin = 600,xmax = 1600,
	ymin = 0,
	ymax = 100,
	xtick = {600,800,...,1600},
	grid = major,
	legend cell align = {left},
      legend style={at={(1,0)},anchor=south east, font = \scriptsize}
	]
	\addplot[
	black,
      only marks,
	mark=triangle*,
	mark repeat = 2,
	mark size = 3,
	]
	table {Final_dats/fig2/RIS_noopt_750.dat};
 \addlegendentry{$L=750 $ m}
 \addplot[
	black,
      only marks,
	mark=*,
	mark repeat = 2,
	mark size = 2,
	]
	table {Final_dats/fig2/RIS_noopt_500.dat};
 \addlegendentry{$L=500 $ m}
  \addplot[
	black,
      only marks,
	mark=square*,
	mark repeat = 2,
	mark size = 2,
	]
	table {Final_dats/fig2/RIS_noopt_250.dat};
 \addlegendentry{$L=250 $ m}
	\addplot[
	black,
       no marks,
	line width = 0.95pt,
	style = densely dashed,
	]
	table {Final_dats/fig2/RIS_noopt_750.dat};
        \addlegendentry{Benchmark}
    \addplot[
	black,
       no marks,
	line width = 0.95pt,
	style = solid,
	]
	table {Final_dats/fig2/RIS_opt_750.dat};
        \addlegendentry{Algorithm 1}
    \addplot[
	black,
      only marks,
	mark=triangle*,
	mark repeat = 2,
	mark size = 3,
	]
	table {Final_dats/fig2/RIS_opt_750.dat};
 \addplot[
	black,
       no marks,
	line width = 0.95pt,
	style = densely dashed,
	]
	table {Final_dats/fig2/RIS_noopt_500.dat};
 \addplot[
	black,
       no marks,
	line width = 0.95pt,
	style = densely dashed,
	]
	table {Final_dats/fig2/RIS_noopt_250.dat};
     \addplot[
	black,
      only marks,
	mark=*,
	mark repeat = 2,
	mark size = 2,
	]
	table {Final_dats/fig2/RIS_opt_500.dat};
     \addplot[
	black,
      only marks,
	mark=square*,
	mark repeat = 2,
	mark size = 2,
	]
	table {Final_dats/fig2/RIS_opt_250.dat};
     \addplot[
	black,
       no marks,
	line width = 0.95pt,
	style = solid,
	]
	table {Final_dats/fig2/RIS_opt_500.dat};
     \addplot[
	black,
       no marks,
	line width = 0.95pt,
	style = solid,
	]
	table {Final_dats/fig2/RIS_opt_250.dat};
	\end{axis}
        \end{tikzpicture}
        \caption{Minimum rate vs the number of reflecting elements $M$ for $\sigma^2=-144$ dB. }
        \label{fig2:RIS}
\end{figure}%

\begin{figure}
        \centering
        \begin{tikzpicture}
 	\begin{axis}[
	width=0.97\linewidth,
	xlabel = {$A_n$},
	ylabel = {$r_\mathrm{min}$ (Mbps)},
	xmin = 2,xmax = 12,
	ymin = 100,
	ymax = 200,
	xtick = {2,4,...,12},
	grid = major,
	legend cell align = {left},
      legend style={at={(1,0)},anchor=south east, font = \scriptsize}
	]
	\addplot[
	black,
      only marks,
	mark=triangle*,
	mark repeat = 1,
	mark size = 3,
	]
	table {Final_dats/fig2/Relay_noopt_750.dat};
 \addlegendentry{$L=750 $ m}
 \addplot[
	black,
      only marks,
	mark=*,
	mark repeat = 1,
	mark size = 2,
	]
	table {Final_dats/fig2/Relay_noopt_500.dat};
 \addlegendentry{$L=500 $ m}
  \addplot[
	black,
      only marks,
	mark=square*,
	mark repeat = 1,
	mark size = 2,
	]
	table {Final_dats/fig2/Relay_noopt_250.dat};
 \addlegendentry{$L=250 $ m}
	\addplot[
	black,
       no marks,
	line width = 0.95pt,
	style = densely dashed,
	]
	table {Final_dats/fig2/Relay_noopt_750.dat};
        \addlegendentry{Benchmark}
    \addplot[
	black,
       no marks,
	line width = 0.95pt,
	style = solid,
	]
	table {Final_dats/fig2/Relay_opt_750.dat};
        \addlegendentry{Algorithm 1}
    \addplot[
	black,
      only marks,
	mark=triangle*,
	mark repeat = 1,
	mark size = 3,
	]
	table {Final_dats/fig2/Relay_opt_750.dat};
 \addplot[
	black,
       no marks,
	line width = 0.95pt,
	style = densely dashed,
	]
	table {Final_dats/fig2/Relay_noopt_500.dat};
 \addplot[
	black,
       no marks,
	line width = 0.95pt,
	style = densely dashed,
	]
	table {Final_dats/fig2/Relay_noopt_250.dat};
     \addplot[
	black,
      only marks,
	mark=*,
	mark repeat = 1,
	mark size = 2,
	]
	table {Final_dats/fig2/Relay_opt_500.dat};
     \addplot[
	black,
      only marks,
	mark=square*,
	mark repeat = 1,
	mark size = 2,
	]
	table {Final_dats/fig2/Relay_opt_250.dat};
     \addplot[
	black,
       no marks,
	line width = 0.95pt,
	style = solid,
	]
	table {Final_dats/fig2/Relay_opt_500.dat};
     \addplot[
	black,
       no marks,
	line width = 0.95pt,
	style = solid,
	]
	table {Final_dats/fig2/Relay_opt_250.dat};
	\end{axis}
        \end{tikzpicture}
    \caption{Minimum rate versus UAV-mounted FDR Antennas for $\sigma^2=-144$ dB }
        \label{fig2:Relay}
\end{figure}
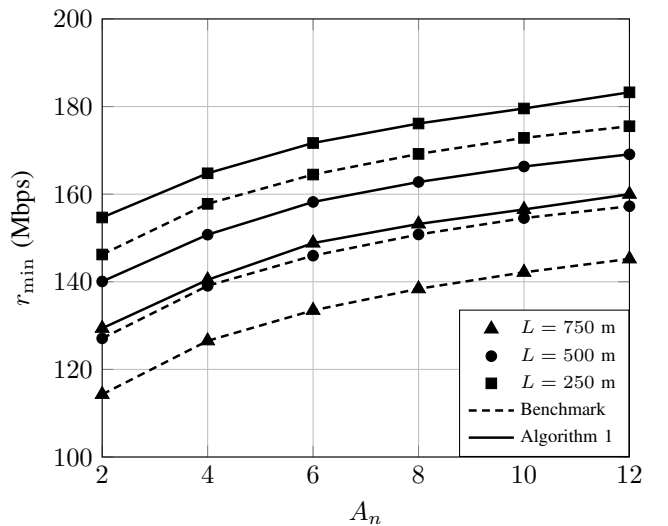

In Fig. 3, we illustrate the relationship between the network's minimum rate $r_{\min}$ and the number of reflecting elements on the UAV-mounted RIS, with the noise power set at $\sigma^2=-144$ dB, for $L=750 $ m, $500 $ m, and $250 $ m, respectively. As it can be observed, $r_{\min}$ increases as $L$ decreases, which is a consequence of the reduced path loss for the GNs. As it can be observed, across all examined $L$ cases, there exists an optimal number of reflecting elements which results from the improved channel gains and the increased energy consumption due to the weight of additional elements. Specifically, adding more reflecting elements increases the rate for each GN logarithmically, while the number of time slots decreases quadratically, in line with (14) and (17) for the case where the UAv speed is equal to zero (i.e., hovering state). Moreover, Fig. 3 shows that for larger $L$ values, a higher number of reflecting elements is optimal, suggesting a strategy to mitigate excess path loss with shorter flight durations. However, in smaller areas, a precise selection of reflecting elements is essential, as fewer elements can achieve enhanced coverage, thus negating the need for further increasing the reflecting elements and avoiding a compromise on the flight duration. Additionally, a key observation is the impact of the optimized user scheduling by Algorithm 1, which significantly reduces the required number of reflecting elements. For instance, under the benchmark scheduling for $L=750 $ m, $500 $ m, and $250 $ m, the optimal numbers of reflecting elements are $1550$, $1500$, and $1100$, respectively, while with optimized user scheduling, these numbers reduce to $1350$, $1150$, and $950$, respectively, emphasizing the advantages of network optimization in determining the optimal RIS size. Finally, in line with Remark 2, the optimal trajectory obtained from Algorithm 1 for the UAV-mounted RIS scenario, considering that $\boldsymbol{q}_{[1]}$ coincides with the BS location, is the hovering trajectory, maximizing the minimum rate through minimized path loss and optimized energy use.

\begin{figure}
	\centering
	\begin{tikzpicture}
	\begin{axis}[
	width=0.97\linewidth,
	xlabel = {$A_n$} ,
	ylabel = {$r_\mathrm{min}$ (Mbps)},
	xmin = 2,xmax = 12,
        ymin=0,
	ymax =13,
	ytick = {0,2,...,12},
 	xtick = {2,4,...,12},
	grid = major,
	grid = major,
	legend image post style={xscale=0.9},
	legend cell align = {left},
        legend style={at={(0,1)},anchor=north west}
	]
	\addplot[
	black,
        only marks,
	mark=triangle,
	mark repeat = 1,
	mark size = 3,
	]
	table {Final_dats/fig3/circ_nopt_outages_relay.dat};
        \addlegendentry{Circle}
    \addplot[
	black,
        only marks,
	mark=o,
	mark repeat = 1,
	mark size = 2,
	]
	table {Final_dats/fig3/rom_nopt_outages_relay.dat};
        \addlegendentry{Rhombus}
    \addplot[
	black,
        only marks,
	mark=square,
	mark repeat = 1,
	mark size = 2,
	]
	table {Final_dats/fig3/spir_nopt_outages_relay.dat};
        \addlegendentry{Spiral}
    \addplot[
	black,
       no marks,
	line width = 0.95pt,
	style = densely dashed,
	]
	table {Final_dats/fig3/circ_nopt_outages_relay.dat};
        \addlegendentry{Benchmark}
    \addplot[
	black,
	mark=diamond*,
	mark repeat = 1,
	mark size = 3.5,
	line width = 0.95pt,
	style = solid,
	]
	table {Final_dats/fig3/circ_opt_outages_relay.dat};
        \addlegendentry{Algorithm 1}
    \addplot[
	black,
       no marks,
	line width = 0.95pt,
	style = densely dashed,
	]
	table {Final_dats/fig3/rom_nopt_outages_relay.dat};
     \addplot[
	black,
       no marks,
	line width = 0.95pt,
	style = densely dashed,
	]
	table {Final_dats/fig3/spir_nopt_outages_relay.dat};
	\end{axis}
	\end{tikzpicture}
	\caption{Minimum rate versus UAV-mounted FDR Antennas for $\sigma^2=-114$ dB and $L=750$ m.}
	\label{fig:Fig4}
\end{figure}
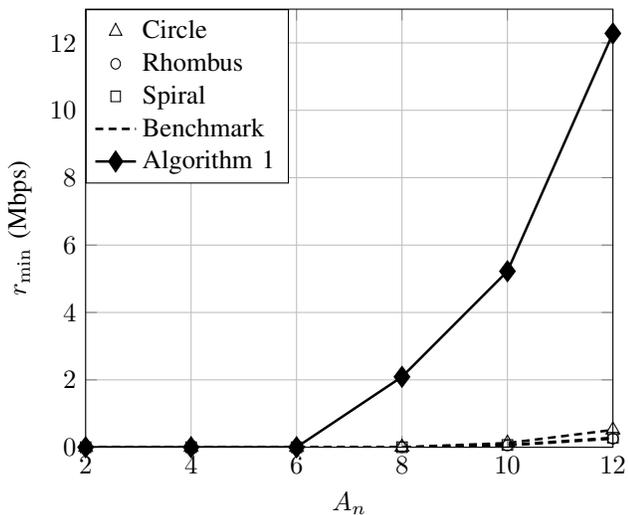

Fig. 4 shows the effect of the number of antennas on the UAV-mounted FDR on $r_{\min}$, for $\sigma^2=-144$ dB and $L=750 $ m, $500 $ m, and $250 $ m, respectively. As it can be seen, decreasing the value of $L$ and increasing the number of FDR antennas enhances $r_{\min}$, with 12 antennas emerging as the optimal number for network performance. Furthermore, similarly to the UAV-mounted RIS scenario, the application of Algorithm 1 further enhances the minimum rate, demonstrating its value on the network performance improvement. Interestingly, under the network parameters in Table II, Algorithm 1 identifies the hovering trajectory as the optimal approach for all the examined cases, which is a notable deviation from the expected optimization of the UAV-mounted FDR’s path loss at intermediate distances between a GN and the BS. Specifically, for the given network parameters, the received SNR for the GNs consistently stays above $\gamma_{\mathrm{thr}}$ across all $L$ values, influencing the trajectory design, as it leads to the conclusion that maximizing the flight duration, thus serving the GNs from the initial UAV position, is more advantageous than moving the UAV to each GN's optimal point, which results in the loss of time slots. In more detail, during the UAV's traversal to these optimal points, there would be instances where the UAV is not optimally positioned to serve any GN, leading to inefficient use of energy and further loss of time slots. Finally, the results from Figs. 3 and 4 indicate that the UAV-mounted FDR outperforms the UAV-mounted RIS for all values of $L$, even with 2 antennas on the FDR. This superior performance of the FDR is attributed to the severe impact of double path loss in the RIS scenario and the FDR's lower energy consumption compared to the RIS, which incurs significant energy costs due to its weight. Additionally, the UAV-mounted FDR's performance is favored as no extra energy is consumed for transitioning, despite potential path loss optimization for each GN. To this end, the provided results underscore the practical implications of UAV system choices on network performance, particularly emphasizing the balance between energy consumption and effective communication, while also highlighting the significant effect of the UAV motors over the energy required for the RIS and FDR operation in the UAV energy consumption.

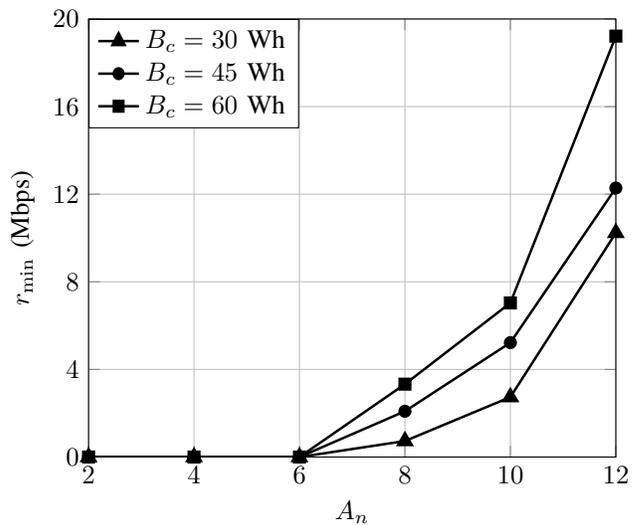
\begin{figure}
	\centering
	\begin{tikzpicture}
	\begin{axis}[
	width=0.97\linewidth,
	xlabel = {$A_n$} ,
	ylabel = {$r_\mathrm{min}$ (Mbps)},
	xmin = 2,xmax = 12,
        ymin=0,
	ymax =20,
	ytick = {0,4,...,20},
 	xtick = {2,4,...,12},
	grid = major,
	grid = major,
	legend image post style={xscale=0.9},
	legend cell align = {left},
        legend style={at={(0,1)},anchor=north west}
	]
	\addplot[
	black,
	mark=triangle*,
	mark repeat = 1,
	mark size = 3,
 	line width = 0.95pt,
	style = solid,
	]
	table {Final_dats/fig5/circ_30Wh.dat};
        \addlegendentry{$B_c=30$ Wh}
    \addplot[
	black,
	mark=*,
	mark repeat = 1,
	mark size = 2,
 	line width = 0.95pt,
	style = solid,
	]
	table {Final_dats/fig5/circ_45Wh.dat};
        \addlegendentry{$B_c=45$ Wh}
    \addplot[
	black,
	mark=square*,
	mark repeat = 1,
	mark size = 2,
 	line width = 0.95pt,
	style = solid,
	]
	table {Final_dats/fig5/circ_60Wh.dat};
        \addlegendentry{$B_c=60$ Wh}
	\end{axis}
	\end{tikzpicture}
	\caption{Minimum rate versus UAV-mounted FDR Antennas for $\sigma^2=-114$ dB, and $L=750$ m for various battery capacities.}
	\label{fig:Fig5}
\end{figure}

Fig. 5 illustrates the trajectory optimization for a UAV-mounted FDR, comparing the solution from Algorithm 1 with the benchmark trajectories for which the benchmark TDMA scheme is applied, in a scenario with an increased noise power of $-114$ dB for an area of $L$ equal to 750 meters. In this scenario, unlike in Fig. 4 where $\sigma^2=-144$ dB and the UAV-mounted FDR remains stationary, the increased noise level results in areas where, if GNs are located within them, the received SNR falls below $\gamma_{\mathrm{thr}}$, UAV movement is necessary to maintain effective communication.  As it can be seen, for $A_n=12$, the circular trajectory achieves a minimum rate of 0.51 Mbps, the rhombus 0.25 Mbps, and the spiral 0.28 Mbps, while the trajectory optimized through Algorithm 1 leads to a significantly higher rate of 12.28 Mbps. Moreover, for all examined trajectories, when the number of $A_n$ is less than 6, the network's minimum rate tends towards zero, due to the increased probability of GNs experiencing outages and the UAV's battery limitations, which are insufficient to enable the UAV to serve all GNs effectively. Therefore, Fig. 5 emphasizes the critical interplay between the number of antennas, UAV battery capacity, and trajectory optimization in ensuring robust network performance, especially in scenarios with challenging communication conditions such as increased noise or $\gamma_{\mathrm{thr}}$ thresholds. Finally, it should be noted that for the UAV-mounted RIS case, if $\sigma^2= -114 dB$ then $r_{\mathrm{min}}=0$ across all the feasible $M$ values, proving again the superiority of UAV-mounted FDR over UAV-mounted RIS in establishing communication link between the BS and the GNs.

\begin{figure*}
    \centering
    \begin{subfigure}{.32\textwidth}
    {	\begin{tikzpicture}
	\begin{axis}[
	height=0.97\linewidth,
	width=0.97\linewidth,
	xlabel = {Length (m)},
	ylabel = {Width (m)},
	xmin = -375,xmax = 375,
	ymin = -375,
	ymax = 375,
	ytick = {-375,-250,...,375},
	xtick = {-375,-250,...,375},
	grid = major,
	legend image post style={xscale=0.9},
	legend cell align = {left},
      legend style={at={(1,0)},anchor=south east,font = \footnotesize}
	]
	\addplot[
	blue,
        only marks,
	mark=triangle*,
	mark repeat = 1,
	mark size = 3,
	]
	table {Final_dats/fig1/aggregators_position.dat};
  	\addlegendentry{GN}
 	\addplot[
	black,
        only marks,
	mark=*,
	mark repeat = 1,
	mark size = 2,
	]
	table {Final_dats/fig1/BS.dat};
 	\addlegendentry{BS}
	\addplot[
	black,
        no marks,
	line width = 0.95pt,
	style = solid,
	]
	table {Final_dats/fig6/relay_outages_best_traj_30Wh.dat};
	\end{axis}
    \end{tikzpicture}}
    \caption{} \label{fig:traj30Wh} 
    \end{subfigure}
    \centering
    \begin{subfigure}{.32\textwidth}{\begin{tikzpicture}
	\begin{axis}[
	height=0.97\linewidth,
	width=0.97\linewidth,
	xlabel = {Length (m)},
	ylabel = {Width (m)},
	xmin = -375,xmax = 375,
	ymin = -375,
	ymax = 375,
	ytick = {-375,-250,...,375},
	xtick = {-375,-250,...,375},
	grid = major,
	legend image post style={xscale=0.9},
	legend cell align = {left},
      legend style={at={(1,0)},anchor=south east,font = \footnotesize}
	]
	\addplot[
	blue,
        only marks,
	mark=triangle*,
	mark repeat = 1,
	mark size = 3,
	]
	table {Final_dats/fig1/aggregators_position.dat};
  	\addlegendentry{GN}
 	\addplot[
	black,
        only marks,
	mark=*,
	mark repeat = 1,
	mark size = 3,
	]
	table {Final_dats/fig1/BS.dat};
 	\addlegendentry{BS}
 	\addplot[
	black,
        no marks,
	line width = 0.95pt,
	style = solid,
	]
	table {Final_dats/fig6/relay_outages_best_traj_45Wh.dat};
	\end{axis}
	\end{tikzpicture}}
    \caption{} \label{fig:traj45Wh} 
    \end{subfigure}
        \begin{subfigure}{.32\textwidth}{\begin{tikzpicture}
	\begin{axis}[
	height=0.97\linewidth,
	width=0.97\linewidth,
	xlabel = {Length (m)},
	ylabel = {Width (m)},
	xmin = -375,xmax = 375,
	ymin = -375,
	ymax = 375,
	ytick = {-375,-250,...,375},
	xtick = {-375,-250,...,375},
	grid = major,
	legend image post style={xscale=0.9},
	legend cell align = {left},
      legend style={at={(1,0)},anchor=south east,font = \footnotesize}
	]
	\addplot[
	blue,
        only marks,
	mark=triangle*,
	mark repeat = 1,
	mark size = 3,
	]
	table {Final_dats/fig1/aggregators_position.dat};
  	\addlegendentry{GN}
 	\addplot[
	black,
        only marks,
	mark=*,
	mark repeat = 1,
	mark size = 2,
	]
	table {Final_dats/fig1/BS.dat};
 	\addlegendentry{BS}
 	\addplot[
	black,
        no marks,
	line width = 0.95pt,
	style = solid,
	]
	table {Final_dats/fig6/relay_outages_best_traj_60Wh.dat};
	\end{axis}
	\end{tikzpicture}}
    \caption{} \label{fig:traj60Wh} 
    \end{subfigure}
    \caption{ Optimal UAV trajectories: (a) $B_c=30$ Wh (b) $B_c=45$ Wh (c) $B_c=60$ Wh.}
    \vspace{-5mm}
\end{figure*}
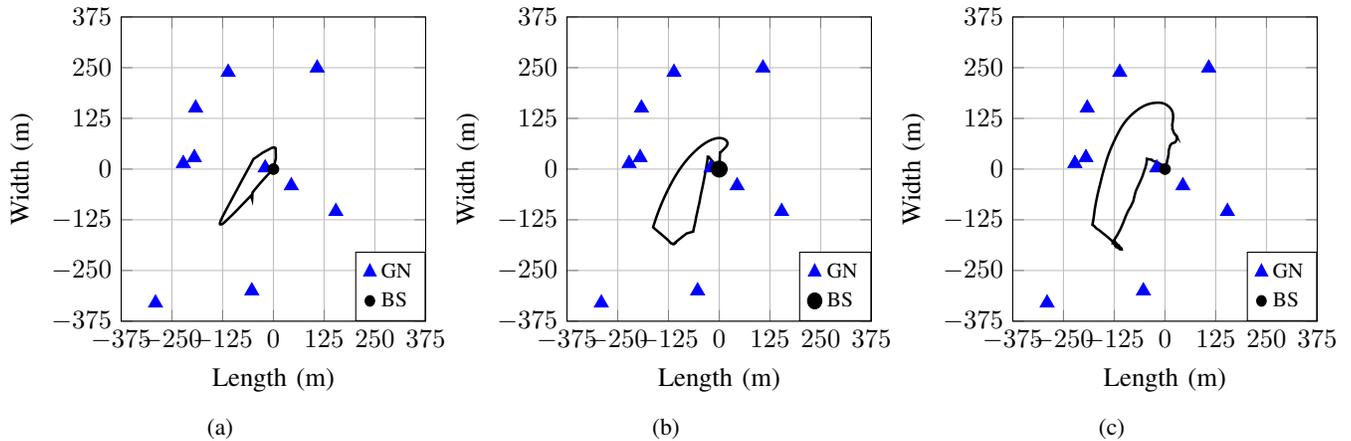

Finally, Fig. 6 depicts how $r_\mathrm{min}$ is affected by the number of antennas $A_n$ of a UAV-mounted FDR across various battery capacities, $B_c$, showcasing how an increase in $B_c$ leads to an improvement in the network's minimum rate. Specifically, this enhancement occurs due to the UAV's ability to allocate more time slots for serving each GN, which not only allows for more extensive service coverage but also enables the UAV to fly to more optimal locations for serving each GN, as the increased number of time slots extends the operational time before the battery depletes. Furthermore, Fig. 6 also indicates that larger battery capacities permit the UAV to execute broader trajectories, optimizing its positioning across the field to serve the GNs more effectively, while smaller battery capacities limit the trajectory size. Complementing this, Fig. 7 illustrates the optimal trajectory for a specific GN setup, which depicts the expansion in the size of the UAV’s trajectory as $B_c$ increases. This expansion reflects the variation in the optimal number of time slots needed for different $B_c$, illustrating the direct relation between battery capacity and the effectiveness of the UAV's flight path in enhancing network performance. Additionally, a notable aspect of these trajectories in Fig. 7, is the formation of a distinct spike in the bottom-left region of the trajectory for all analyzed battery capacities, which is a consequence of the strategic positioning needed to address the unfavorable location of the bottom-left GNs within the rectangular area. This contrasts with the right-hand side of the rectangular area, where the GNs are fewer in number, leading to a different trajectory pattern that does not necessitate such pronounced adjustments. To this end, Fig. 6 and Fig. 7 underscore the effect of the battery capacity on trajectory optimization, while emphasizing their significance in ensuring robust network performance, particularly in scenarios with demanding communication conditions.


\section{Conclusions}\label{section:conc}
In this work, a thorough comparison between UAV-mounted RIS and UAV-mounted FDRs was performed, highlighting their distinct energy consumption patterns and underscoring the critical role of energy-aware design in UAV-based communication networks. Our results elucidated an optimal number of reflective elements for the RIS and antennas for the FDR, precisely tailored to the studied scenario, with our optimization algorithm suggesting a possible reduction in the optimal number of reflective elements. The analysis showed that increasing the number of reflective elements for the RIS logarithmically improves the rate for each GN, but also increases the energy consumption due to the added weight, which represents a trade-off between channel gain and energy efficiency. Despite the nearly passive nature of the RIS, its significant weight, coupled with the inherent double path loss, challenges its applicability in the considered scenario. In contrast, the FDR was consistently identified as a more efficient solution, optimizing network fairness through its favorable path loss without significantly impacting the UAV's energy consumption. Furthermore, the primary energy consumption factor was identified as the UAV's motors, highlighting the importance of lightweight design, especially for UAV-mounted RIS systems. Finally, it was shown that the UAV's battery energy significantly influences the optimal trajectory, forcing the UAV to move only when necessary to minimize energy consumption, thus emphasizing the importance of energy awareness in the strategic operation of UAV-assisted networks. Therefore, our study lays the foundation for future exploration of advanced RIS technologies, e.g., metasurface-based RIS and non-orthogonal multiple access schemes, which promise to unravel complex dynamics and further improve the efficiency and sustainability of UAV-assisted communication networks.


\bibliographystyle{IEEEtran}
\bibliography{Bibliography}

\begin{thebibliography}{10}
\providecommand{\url}[1]{#1}
\csname url@samestyle\endcsname
\providecommand{\newblock}{\relax}
\providecommand{\bibinfo}[2]{#2}
\providecommand{\BIBentrySTDinterwordspacing}{\spaceskip=0pt\relax}
\providecommand{\BIBentryALTinterwordstretchfactor}{4}
\providecommand{\BIBentryALTinterwordspacing}{\spaceskip=\fontdimen2\font plus
\BIBentryALTinterwordstretchfactor\fontdimen3\font minus
  \fontdimen4\font\relax}
\providecommand{\BIBforeignlanguage}[2]{{%
\expandafter\ifx\csname l@#1\endcsname\relax
\typeout{** WARNING: IEEEtran.bst: No hyphenation pattern has been}%
\typeout{** loaded for the language `#1'. Using the pattern for}%
\typeout{** the default language instead.}%
\else
\language=\csname l@#1\endcsname
\fi
#2}}
\providecommand{\BIBdecl}{\relax}
\BIBdecl

\bibitem{zorzi2021}
M.~Giordani and M.~Zorzi, ``{Non-terrestrial networks in the 6G era: Challenges
  and opportunities},'' \emph{IEEE Netw.}, vol.~35, no.~2, pp. 244--251, 2021.

\bibitem{Mekikis2023}
P.-V. Mekikis, P.~S. Bouzinis, N.~A. Mitsiou, S.~A. Tegos, D.~Tyrovolas, V.~K.
  Papanikolaou, and G.~K. Karagiannidis, ``{Enabling wireless-powered IoT
  through incentive-based UAV swarm orchestration},'' \emph{IEEE Open J.
  Commun. Soc}, vol.~4, pp. 2548--2560, 2023.

\bibitem{Matracia2023}
M.~Matracia, M.~A. Kishk, and M.-S. Alouini, ``{Comparing aerial-RIS- and
  aerial-base-station-aided post-disaster cellular networks},'' \emph{IEEE Open
  J. Veh. Technol.}, pp. 1--15, 2023.

\bibitem{ruizhang2017}
Y.~Zeng and R.~Zhang, ``{Energy-efficient UAV communication with trajectory
  optimization},'' \emph{IEEE Trans. Veh. Technol.}, vol.~16, no.~6, pp.
  3747--3760, 2017.

\bibitem{YongZeng2019}
Y.~Zeng, J.~Xu, and R.~Zhang, ``{Energy minimization for wireless communication
  with rotary-wing UAV},'' \emph{IEEE Trans. Wirel. Commun.}, vol.~18, no.~4,
  pp. 2329--2345, 2019.

\bibitem{hanzo2016}
Z.~Zhang, K.~Long, A.~V. Vasilakos, and L.~Hanzo, ``{Full-duplex wireless
  communications: Challenges, solutions, and future research directions},''
  \emph{Proc. IEEE}, vol. 104, no.~7, pp. 1369--1409, 2016.

\bibitem{shende2018}
N.~V. Shende, O.~G\"urb\"uz, and E.~Erkip, ``{Half-duplex or full-duplex
  communications: Degrees of freedom analysis under self-interference},''
  \emph{IEEE Trans. Wirel. Commun.}, vol.~17, no.~2, pp. 1081--1093, 2018.

\bibitem{yuanweiliu2021}
Y.~Liu, X.~Liu, X.~Mu, T.~Hou, J.~Xu, M.~Di~Renzo, and N.~Al-Dhahir,
  ``{Reconfigurable intelligent surfaces: Principles and opportunities},''
  \emph{IEEE Commun. Surv. Tutor.}, vol.~23, no.~3, pp. 1546--1577, 2021.

\bibitem{linzhang2017}
L.~Zhang, J.~Liu, M.~Xiao, G.~Wu, Y.-C. Liang, and S.~Li, ``{Performance
  analysis and optimization in downlink NOMA systems with cooperative
  full-duplex relaying},'' \emph{IEEE J. Sel. Areas Commun.}, vol.~35, no.~10,
  pp. 2398--2412, 2017.

\bibitem{liaskos2018}
C.~Liaskos, S.~Nie, A.~Tsioliaridou, A.~Pitsillides, S.~Ioannidis, and
  I.~Akyildiz, ``A new wireless communication paradigm through
  software-controlled metasurfaces,'' \emph{IEEE Commun. Mag.}, vol.~56, no.~9,
  pp. 162--169, 2018.

\bibitem{basar2019}
E.~Basar, M.~Di~Renzo, J.~De~Rosny, M.~Debbah, M.-S. Alouini, and R.~Zhang,
  ``{Wireless communications through reconfigurable intelligent surfaces},''
  \emph{IEEE Access}, vol.~7, pp. 116\,753--116\,773, 2019.

\bibitem{direnzo2020}
M.~Di~Renzo, A.~Zappone, M.~Debbah, M.-S. Alouini, C.~Yuen, J.~de~Rosny, and
  S.~Tretyakov, ``Smart radio environments empowered by reconfigurable
  intelligent surfaces: How it works, state of research, and the road ahead,''
  \emph{IEEE J. Sel. Areas Commun.}, vol.~38, no.~11, pp. 2450--2525, 2020.

\bibitem{gangliu2015}
G.~Liu, F.~R. Yu, H.~Ji, V.~C.~M. Leung, and X.~Li, ``In-band full-duplex
  relaying: A survey, research issues and challenges,'' \emph{IEEE Commun.
  Surv. Tutor.}, vol.~17, no.~2, pp. 500--524, 2015.

\bibitem{yihong2022}
M.~Tatar~Mamaghani and Y.~Hong, ``Aerial intelligent reflecting surface-enabled
  terahertz covert communications in beyond-{5G} internet of things,''
  \emph{IEEE Internet Things J.}, vol.~9, no.~19, pp. 19\,012--19\,033, 2022.

\bibitem{haas2021}
T.~N. Do, G.~Kaddoum, T.~L. Nguyen, D.~B. da~Costa, and Z.~J. Haas, ``{Aerial
  reconfigurable intelligent surface-aided wireless communication systems},''
  in \emph{Proc. IEEE 32nd Annual International Symposium on Personal, Indoor
  and Mobile Radio Communications (PIMRC)}, 2021, pp. 525--530.

\bibitem{trung2021}
Y.~Li, C.~Yin, T.~Do-Duy, A.~Masaracchia, and T.~Q. Duong, ``{Aerial
  reconfigurable intelligent surface-enabled URLLC UAV systems},'' \emph{IEEE
  Access}, vol.~9, pp. 140\,248--140\,257, 2021.

\bibitem{chatzinotas2022}
S.~Solanki, S.~Gautam, S.~K. Sharma, and S.~Chatzinotas, ``{Ambient backscatter
  assisted co-existence in aerial-IRS wireless networks},'' \emph{IEEE Open J.
  Commun. Soc.}, vol.~3, pp. 608--621, 2022.

\bibitem{pitilakis2023}
A.~Pitilakis, D.~Tyrovolas, P.-V. Mekikis, S.~A. Tegos, A.~Papadopoulos,
  A.~Tsioliaridou, O.~Tsilipakos, D.~Manessis, S.~Ioannidis, N.~V. Kantartzis,
  I.~F. Akyildiz, and C.~K. Liaskos, ``On the mobility effect in {UAV}-mounted
  absorbing metasurfaces: A theoretical and experimental study,'' \emph{IEEE
  Access}, vol.~11, pp. 79\,777--79\,792, 2023.

\bibitem{haibomei2022}
H.~Mei, K.~Yang, Q.~Liu, and K.~Wang, ``{3D-trajectory and phase-shift design
  for RIS-assisted UAV systems using deep reinforcement learning},'' \emph{IEEE
  Trans. Veh. Technol.}, vol.~71, no.~3, pp. 3020--3029, 2022.

\bibitem{huilong2020}
H.~Long, M.~Chen, Z.~Yang, B.~Wang, Z.~Li, X.~Yun, and M.~Shikh-Bahaei,
  ``{Reflections in the sky: Joint trajectory and passive beamforming design
  for secure UAV networks with reconfigurable intelligent surface},'' 2020.

\bibitem{binduo2023}
B.~Duo, M.~He, Q.~Wu, and Z.~Zhang, ``{Joint dual-UAV trajectory and RIS design
  for ARIS-assisted aerial computing in IoT},'' \emph{IEEE Internet Things J.},
  pp. 1--1, 2023.

\bibitem{menghua2018}
M.~Hua, Y.~Wang, Z.~Zhang, C.~Li, Y.~Huang, and L.~Yang, ``Outage probability
  minimization for low-altitude {UAV}-enabled full-duplex mobile relaying
  systems,'' \emph{China Communications}, vol.~15, no.~5, pp. 9--24, 2018.

\bibitem{depaiva2021}
D.~De~Paiva~Mucin, D.~P.~M. Osorio, and E.~E.~B. Olivo, ``Wireless-powered
  full-duplex {UAV} relay networks over {FTR} channels,'' \emph{IEEE Open J.
  Commun. Soc.}, vol.~2, pp. 2205--2218, 2021.

\bibitem{lipengzhu2020}
L.~Zhu, J.~Zhang, Z.~Xiao, X.~Cao, X.-G. Xia, and R.~Schober, ``Millimeter-wave
  full-duplex uav relay: Joint positioning, beamforming, and power control,''
  \emph{IEEE J. Sel. Areas Commun.}, vol.~38, no.~9, pp. 2057--2073, 2020.

\bibitem{binduo2020}
B.~Duo, Q.~Wu, X.~Yuan, and R.~Zhang, ``Energy efficiency maximization for
  full-duplex {UAV} secrecy communication,'' \emph{IEEE Trans. Veh. Technol.},
  vol.~69, no.~4, pp. 4590--4595, 2020.

\bibitem{bingli2021}
B.~Li, S.~Zhao, R.~Zhang, and L.~Yang, ``Full-duplex {UAV} relaying for
  multiple user pairs,'' \emph{IEEE Internet Things J.}, vol.~8, no.~6, pp.
  4657--4667, 2021.

\bibitem{WeiWang2022}
W.~Wang, N.~Qi, L.~Jia, C.~Li, T.~A. Tsiftsis, and M.~Wang, ``{Energy-efficient
  UAV-relaying 5G/6G spectrum sharing networks: Interference coordination with
  power management and trajectory design},'' \emph{IEEE Open J. Commun. Soc},
  vol.~3, pp. 1672--1687, 2022.

\bibitem{tyrovolas2023}
D.~Tyrovolas, P.-V. Mekikis, S.~A. Tegos, P.~D. Diamantoulakis, C.~K. Liaskos,
  and G.~K. Karagiannidis, ``{Energy-aware design of UAV-mounted RIS networks
  for IoT data collection},'' \emph{IEEE Trans. Commun.}, vol.~71, no.~2, pp.
  1168--1178, 2023.

\bibitem{shaikh2021}
M.~H.~N. Shaikh, V.~A. Bohara, A.~Srivastava, and G.~Ghatak, ``{Intelligent
  reflecting surfaces versus full-duplex relaying: Performance comparison for
  non-ideal transmitter case},'' in \emph{Proc. IEEE 32nd Annual International
  Symposium on Personal, Indoor and Mobile Radio Communications (PIMRC)}, 2021,
  pp. 513--518.

\bibitem{yue2023}
Y.~Xiao, D.~Tyrovolas, S.~A. Tegos, P.~D. Diamantoulakis, Z.~Ma, L.~Hao, and
  G.~K. Karagiannidis, ``{Solar powered UAV-mounted RIS networks},'' \emph{IEEE
  Commun. Lett.}, vol.~27, no.~6, pp. 1565--1569, 2023.

\bibitem{mdpi2023}
\BIBentryALTinterwordspacing
C.~Y. Goh, C.~Y. Leow, and R.~Nordin, ``Energy efficiency of unmanned aerial
  vehicle with reconfigurable intelligent surfaces: A comparative study,''
  \emph{Drones}, vol.~7, no.~2, 2023. [Online]. Available:
  \url{https://www.mdpi.com/2504-446X/7/2/98}
\BIBentrySTDinterwordspacing

\bibitem{mekikis2019}
P.-V. Mekikis and A.~Antonopoulos, ``Breaking the boundaries of aerial networks
  with charging stations,'' in \emph{Proc. IEEE International Conference on
  Communications (ICC)}, 2019, pp. 1--6.

\bibitem{talebi2008}
A.~Talebi and W.~A. Krzymien, ``Multiple-antenna multiple-relay cooperative
  communication system with beamforming,'' in \emph{Proc. IEEE Vehicular
  Technology Conference}, 2008, pp. 2350--2354.

\bibitem{jianhui2022}
J.~Ma, C.~Huang, and Q.~Li, ``Energy efficiency of full- and half-duplex
  decode-and-forward relay channels,'' \emph{IEEE Internet Things J.}, vol.~9,
  no.~12, pp. 9730--9748, 2022.

\end{thebibliography}

\end{document}